\documentclass[aoas,preprint]{imsart}

\RequirePackage{amsthm,amsmath,amsfonts,amssymb}
\RequirePackage[authoryear]{natbib}
\RequirePackage[colorlinks,citecolor=blue,urlcolor=blue]{hyperref}
\RequirePackage{graphicx}
\usepackage{listings}
\usepackage{placeins}
\usepackage{longtable}
\usepackage{comment}

\startlocaldefs

\endlocaldefs

\begin{document}

\begin{frontmatter}

\title{Estimating animal utilization distributions from multiple data types: a joint spatio-temporal point process framework}
\runtitle{A framework for estimating UDs from static and mobile observers}

\begin{aug}

\author[A]{\fnms{Joe} \snm{Watson} \ead[label=e1]{joe.watson@stat.ubc.ca}},
\author[B,C]{\fnms{Ruth} \snm{Joy}\ead[label=e3]{}},
\author[B]{\fnms{Dominic} \snm{Tollit}\ead[label=e2]{}},
\author[D]{\fnms{Sheila J} \snm{Thornton}\ead[label=e4]{}},
\and
\author[A,E]{\fnms{Marie} \snm{Auger-M\'eth\'e}\ead[label=e5]{}}


\address[A]{Department of Statistics, University of British Columbia, Vancouver, Canada, \printead{e1}}
\address[B]{SMRU Consulting, Vancouver, Canada \printead{e2}}
\address[C]{School of Environmental Science, Simon Fraser University, Burnaby, Canada \printead{e3}}
\address[D]{Fisheries and Oceans Canada, Pacific Science Enterprise Centre,
West Vancouver,
Canada \printead{e4}}
\address[E]{Institute for the Oceans and Fisheries, University of British Columbia, Vancouver, Canada \printead{e5}}
\end{aug}

\begin{abstract} Models of the spatial distribution of animals provide useful tools to help ecologists quantify species-environment relationships, and they are increasingly being used to help determine the impacts of climate and habitat changes on species. While high-quality survey-style data with known effort are sometimes available, often researchers have multiple datasets of varying quality and type. In particular, collections of sightings made by citizen scientists are becoming increasingly common, with no information typically provided on their observer effort. Many standard modelling approaches ignore observer effort completely, which can severely bias estimates of an animal's distribution. Combining sightings data from observers who followed different protocols is challenging. Any differences in observer skill, spatial effort, and the detectability of the animals across space all need to be accounted for. To achieve this, we build upon the recent advancements made in integrative species distribution models and present a novel marked spatio-temporal point process framework for estimating the utilization distribution (UD) of the individuals of a highly mobile species. We show that in certain settings, we can also use the framework to combine the UDs from the sampled individuals to estimate the species' distribution. We combine the empirical results from a simulation study with the implications outlined in a causal directed acyclic graph to identify the necessary assumptions required for our framework to control for observer effort when it is unknown. We then apply our framework to combine multiple datasets collected on the endangered Southern Resident Killer Whales, to estimate their monthly effort-corrected space-use. 
\end{abstract}


\end{frontmatter}











\section{Introduction}

Accurate knowledge of the spatio-temporal distribution of animals is vital for understanding the effects of climate and habitat changes on species and for governments to implement successful management policies. In particular, ecologists are interested in estimating the space use of individual animals to inform a wide range of questions. For example, such estimates can be used to quantify the location and extent of habitats required for a species' conservation strategy \citep{fleming2015rigorous}. Estimates of space use can be linked to environmental covariates within an occupancy model to explain resource use \citep{mordecai2011addressing}. These estimates are also used within a spatial capture-recapture framework for estimating home range centers and population abundance \citep{royle2011spatial}. The space use of an individual is often referred to as the utilization distribution (UD). The UD of an individual defines the probability density that an individual is found at a given point in space and time \citep{lele2013selection}. Estimating individual UDs is often complicated by complex animal movement and observer processes, both of which must be considered within an analysis \citep{royle2011spatial}. Accounting for observers with unknown, but estimable, effort remains an open problem for UDs.

Ecologists use many methods for estimating UDs from different types of individually-identified sightings. In particular, many methods exist for animal tracking datasets where the locations of individuals are followed through time with devices such as GPS tags. Simple methods include geometric techniques such as the minimum convex polygon \citep{fieberg2012could}, and statistical smoothers such as kernel density estimators \citep{worton1989kernel}. Methods have also been developed to account for the autocorrelations due to animal movement \citep{fleming2015rigorous}, but these assume that the sightings were made without observer bias. For tracking data collected at high temporal frequency, resource selection function models can be used to estimate UDs \citep{johnson2013estimating}. Fewer methods exist for sightings (or captures) data made at a set of discrete locations (e.g. camera traps) or by a group of mobile observers with recorded locations. Such datasets are commonly collected \citep{whoriskey2019current, hussey2015aquatic}. In these settings, spatial capture-recapture models can be used to estimate UDs throughout the study region \citep{royle2011spatial}. However, these methods typically assume that the UD of each individual follows a bivariate normal distribution centered at a unique, latent, home range center. This assumed form of the UD may be overly simplistic when large quantities of data are available for each individual. We propose a framework to model complex individual UDs that accounts for the observer efforts from both mobile and static observers and models the relationships between the individuals' space use and environmental characteristics.

A similar but distinct objective to estimating the UDs of individuals is estimating the distribution of a species. Species distribution models (SDM's hereafter) are tools for predicting the density of a species and for relating it to measured environmental covariates. SDMs have been the focus of statistical research for decades \citep{elith2007predicting, fithian2015bias}. In particular, much work has been done on how to use point process methods to develop SDMs that can combine data of varying type (see \citet{miller2019recent} for a recent review), account for observer processes and biases \citep{koshkina2017integrated}, and include spatio-temporal random effects to capture additional autocorrelations \citep{yuan2017point}. Furthermore, point processes are scale-invariant and do not encounter the `pseudo-absence' problem faced by competing methods \citep{warton2010poisson}. Thus, point processes have emerged as the most promising integrative model framework for jointly modeling data of varying types and quality \citep[e.g.][]{renner2015point, giraud2016capitalizing}. By sharing parameters and latent effects between the likelihoods of each data type within a joint model \citep[e.g.][]{giraud2016capitalizing, bedrinana2018integrating}, strength can be borrowed and hence a greater precision in conclusions and improved management policies can be attained \citep{fithian2015bias}. This approach can be made more robust for datasets of especially poor quality, by allowing only a correlative relationship to exist between their models and the latent effects defining the SDM \citep{pacifici2017integrating}.

In this paper, we show how to adapt these recent point process frameworks for use with UDs. In particular, we assume that the individuals' UDs are stationary within well-defined time periods and that we can subset the sightings data to obtain approximately independent snapshots of the UDs. We then argue that these recent developments make point processes an ideal basis to create a framework for modeling individual UDs in data-rich settings. In particular, using point processes for UDs allows us to combine numerous datasets following different complex data collection protocols \citep{miller2019recent}. For example, presence-only sightings that lack any records of locations where sightings were not made (i.e. absences) may be combined with high quality presence-absence data (see \citet{hefley2016hierarchical, miller2019recent} and references within). Accounting for differences in protocols is crucial. Some protocols may focus observer efforts in regions where the density of the species under study is highest. This preferential sampling can lead to positively biased estimates of species density, abundance, and UDs \citep{pennino2019accounting,watson2018general}.  

With the data appropriately subsetted, any autocorrelation between the sightings that could bias inference is removed \citep{johnson2013estimating}. Yet, there remain two hurdles that we must overcome to apply point process methods to UD estimation. First, we must adapt the point process framework to the setting where individuals can be identified at numerous locations through time. Typical SDM analyses often assume that the locations of individuals remain fixed throughout the sampling time \citep{koshkina2017integrated, giraud2016capitalizing}. Second, we must generalise the point process framework to allow for combinations of mobile and static observers with highly complex, and potentially unknown (but estimable) effort. Well-defined discrete sampling `sites' are often assumed to exist \citep{giraud2016capitalizing}. However, in many cases, observer effort is continuous in both space and time. In continuous time, unless strict distance sampling protocols with high observer speed are followed, animal movement can bias estimates of absolute intensity \citep{glennie2015effect, yuan2017point}. Explicitly modelling the animal movement model within the observer's sampling process to account for this can prove challenging \citep{glennie2020incorporating}. This is exacerbated when information on the sampling protocols is unavailable, as is often the case. 

Taking the above concerns into account, we create a flexible framework for estimating the UDs of individuals in data-rich settings. We relax many of the stringent assumptions and requirements made previously, creating a general approach for estimating the spatio-temporal distributions of individual animals. Data from combinations of mobile and static observers, with differing skill levels, and following differing protocols may be combined. Locations of observers may be known (e.g GPS), or unknown. In the latter, either a highly informative set of covariates or an emulator must be available that can adequately describe the observers' efforts. In either case, observer effort can be controlled for, with any uncertainties in effort propagated through to resulting inference. Our approach circumvents the modeling of an animal's movement process by focusing on relative intensity values when estimating UDs. This approach largely avoid the bias discussed by \citet{glennie2020incorporating} for estimating absolute intensity values. We demonstrate this claim through a simulation study and show that large improvements in the predictions of UDs are possible using our approach in settings where the degree of spatial heterogeneity in observer effort is high. 

Using our approach, statistical inference can be made at both an individual level, and/or a population level. We show that population inference is trivial in settings where sightings data of each individual in the population is available. When only a subset of the population is observed, the sampled subpopulation must be representative of the population for extrapolation to be accurate. To adapt the point process framework for use with repeated sightings of individuals, we redefine the intensity surface being modeled as an encounter rate instead of as an expected number of individuals per unit area. Including random fields and random effects can model the additional spatio-temporal correlation induced by unmeasured spatially-smooth covariates and/or biological processes \citep{pacifici2017integrating, yuan2017point}.  Since this approach is subsumed under a log-Gaussian Cox process (LGCP) framework \citep{chakraborty2011point}, implementing this idea is made especially easy using the R package inlabru \citep{R,inlabru}, enabling researchers to adapt this framework for use in their applications.

The paper is structured as follows. First, we present our motivating problem: estimating the spatio-temporal distribution of the Southern Resident Killer Whale between May and October. This species is of special conservation concern. Next, we define the types of encounter events assumed to generate the data. Then, we introduce marked LGCPs as the preferred tool for modeling the encounters. We discuss their properties and define the new intensity surface to be modeled in terms of encounter rates and effort intensities. After presenting our modeling framework, we demonstrate its properties in a thorough simulation study, before adapting it to our motivating problem. Finally, we design easily-interpretable maps to display the UDs of the whales across the months. We then aggregate these to provide inference on both a pod (i.e.\ group) level and a population level. Computer code using the inlabru package \citep{inlabru} is provided to enable researchers to adapt this framework for their applications.

\section{Motivating Problem}

\subsection{An introduction to the problem}

The Southern Resident Killer Whale (SRKW) population is listed as Endangered under both the Canadian Species at Risk Act (\citeauthor{DFO_web}) and the United States Endangered Species Act (\citeauthor{NOAA_ESA_web}) because of their small population size, low reproductive rate, the existence of a variety of anthropogenic threats, and prey availability. The range of the SRKW extends from southeastern Alaska to central California; however, between May - September, all three pods of these whales frequent the waters of both Canada and the United States, concentrating in the Salish Sea and Swiftsure bank (\citeauthor{DFO_web}). We extend our study region beyond these areas (see Fig \ref{fig:WWandDFOsightings}). 

The development of successful and effective policies to help protect the SRKW requires accurate, high-resolution knowledge on how their space use evolves across the calendar year. The SRKW are highly social animals, spending the majority of their time in three well-defined groups called the J, K and L pods \citep{ford1996killer}. Hereafter, we consider pods as individuals for the purposes of modeling. Inter-pod variation in the summer space-use exists \citep{hauser2007summer}. Due to the differing characteristics of the three pods, additional knowledge of space-use on a pod-level may help to improve the effectiveness of management decisions. While SRKWs are known to favour the inshore waters of Washington State and British Columbia in the summer months \citep{ford2017habitats, hauser2006evaluating}, precise knowledge surrounding their space use across the months is lacking, as is precise knowledge surrounding the differences between the pods \citep{hauser2007summer}. 

\subsection{The data available}

Multiple sources of SRKW sightings are available, including GPS-tracked focal follows of targeted individuals by citizen scientists, presence-only sightings from commercial whale-watching vessels, and opportunistic sightings reported by the public. We judge two of these data sources to be most suited for use in an effort-corrected analysis as we are able to estimate its observer effort and we are confident that the source could accurately differentiate between
the three SRKW pods. 

\begin{figure}[ht!]
    \centering
    \includegraphics[scale=0.3]{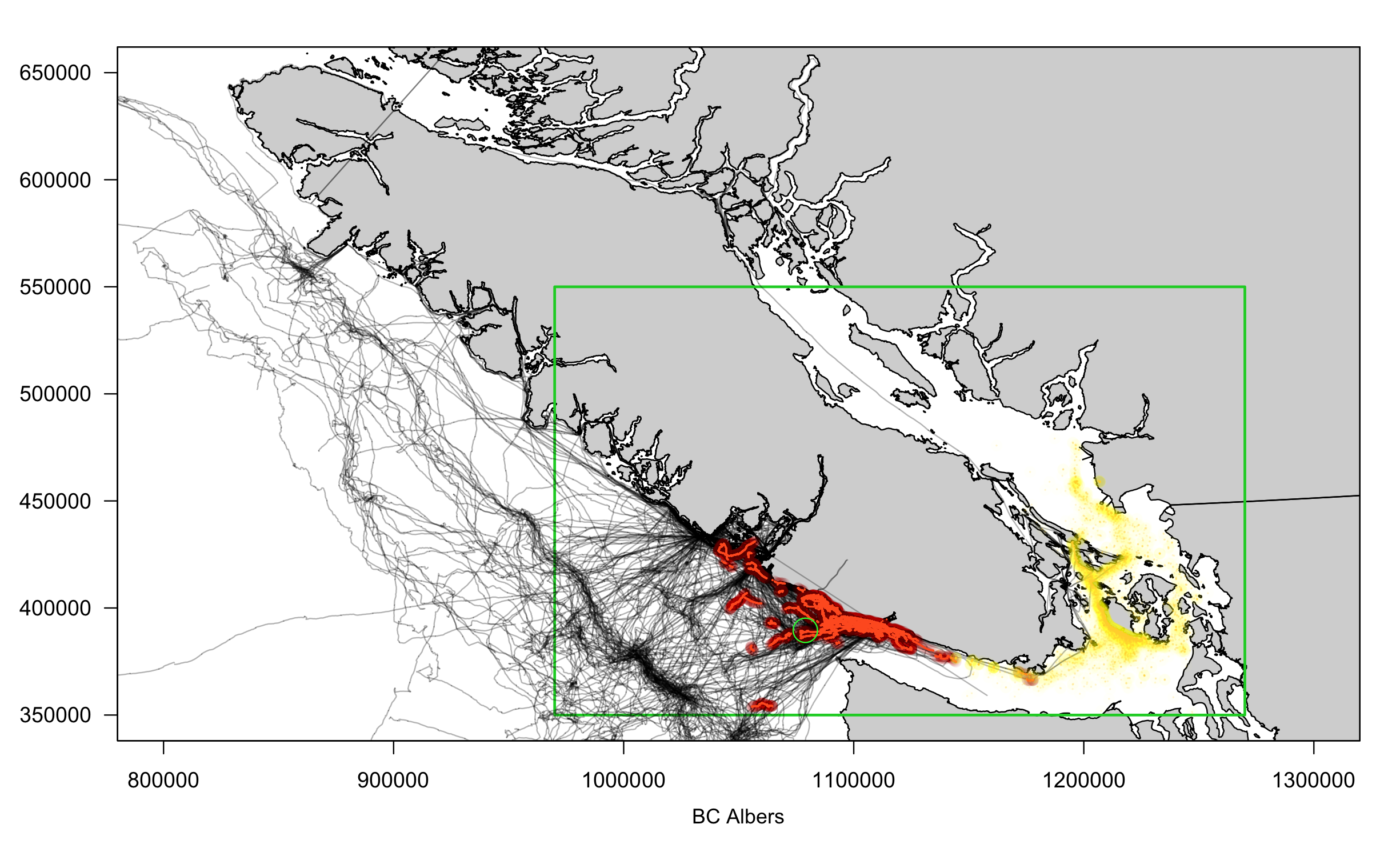}
    \caption{A plot showing our area of interest $\Omega$ in green, with the GPS tracklines of the DFO survey effort displayed as black lines. All DFO survey sightings are shown as a red overlay on top of the effort. All sightings from the OM and BCCSN datasets are shown in yellow. Shown are sightings and tracklines from May - October 2009 - 2016. The green circle roughly locates the Swiftsure Bank, with the waters due East and North of this representing the Salish Sea. The BC Albers projection shown is in units of metres.}
    \label{fig:WWandDFOsightings}
\end{figure}

The first source of data we use are presence-absence data collected through a project funded by the Department of Fisheries and Oceans Canada. The data were collected from a mobile vessel on the west side of the Strait of Juan de Fuca fitted with a GPS tracker between 2009 and 2016. The GPS coordinates of the vessel were recorded at a high frequency, and all sightings of SRKWs were reported, along with the pod (see Fig \ref{fig:WWandDFOsightings}). The marine mammal observer on this vessel is an expert on the SRKW, and thus, the reported pod classification can be considered accurate. The motivations of the Captain's data collection varied from year to year, and hence, the spatial distribution of their observer effort varied. We refer to this dataset as DFO hereafter. 

The second reliable data source we use are the presence-only SRKW sightings reported by the whale-watch industry between 2009 - 2016 and collected by two organisations. The B.C. Cetacean Sightings Network (BCCSN) \citep{BCCSN_web} and The OrcaMaster (OM) [The Whale Museum] \citep{OM} datasets both contain sightings from a vast range of observer types, but we exclusively model the whale-watch sightings for three reasons. First, the whale-watch operators have a high degree of expertise on the SRKW, with vessels typically having a biologist or other expert onboard. This results in accurate pod classifications. Second, the whale-watch companies are known to share the locations of the sighted SRKW with each other. Thus, our dataset likely contains the majority of whale-watch sightings that were made, not just the subset of those made by the operators who report to the databases. Third, a vast amount of data has been collected on the activities of the whale-watching industry operating in the area. This enables us to estimate the observer effort from these companies with a high degree of accuracy and precision. We refer to this combined dataset as WW hereafter. 

The majority of the data we obtained on the activities of the whale-watch industry came from the Soundwatch Boater Education Program (Soundwatch hereafter). Soundwatch is a vessel monitoring and public education outreach program that systematically monitors vessel activities around cetaceans during the
whale-watch season (May - September) in the Haro Strait Region of the Salish Sea \citep{seely2017soundwatch}. Since 2004, Soundwatch has been using data collection protocols established in partnership with NOAA, DFO, and the Canadian Straitwatch Program. This includes detailed accounts of vessel types, whale-watch vessel numbers, and whale-watch vessel activities.

\subsection{Previous work estimating the space use of SRKW}

Previous work estimating the summer space use of SRKW has greatly assisted the development of critical habitat regions and has helped inform successful management initiatives. In the past 15 years, two pieces of published research tackled the problem from different angles. \citet{hauser2007summer} estimated the summer space use of the SRKW within the Salish Sea's inshore waters of Washington and British Columbia. The authors assumed a constant observer effort across the months from the whale-watch operators within their study region, based on results from a field study \citep{hauser2006evaluating}. Pod-specific core areas were identified and SRKW hotspots were clearly displayed. However, their study region was smaller than ours.

Most recently, \citet{OM} estimated an effort-corrected map of SRKW summer space use. This expanded on the work of \citet{hauser2007summer}, by estimating the space use across a larger area than theirs, as well as by incorporating the heterogeneous observer effort in their modeling directly. Regions of `high' effort-adjusted whale density were identified and clearly presented in detailed plots. To reduce the impact of autocorrelation from the whales' movements on the analysis, they defined `whale days' as their target metric. They defined a whale day
to be any day where a SRKW was reported, regardless of the number of times they were reported on that day. A smaller study region relative to ours was studied, and no environmental covariates were used. Estimation of observer effort followed previous unpublished research from the Vancouver Aquarium (\citet{VanAq}, \textit{pers comm}).

\subsection{Goals of the analysis}

We aim to build upon the previous research and build a model for estimating high-resolution, effort-corrected, and temporally-changing SRKW space use across the summer months (May - October). For each pod, their UD will provide the probability that they occupy a specific region at any given instant in time within each month. Under the assumption that the study region properly captures the full spatial extent of the UDs, we can then estimate the spatial distribution of the SRKW for each month. All estimates will be conditioned upon detailed estimates of observer effort from the whale-watch companies, combined with GPS tracks from the DFO dataset. Unlike previous attempts, our model will use multiple environmental covariates such as sea-surface temperature and various measures of primary productivity (e.g. chlorophyll-A) to improve the accuracy of predicted maps. 

By turning to statistical/probabilistic modeling, we will attempt to account for all sources of uncertainties, including the uncertainties associated with our estimates of the observer effort from the whale-watch vessels. This has not previously been done. Finally, we will demonstrate how the methodology allows for the creation of maps that simultaneously display regions of high SRKW intensity for each month, along with their corresponding uncertainties. We display these for the month of May for exhibition. Identifying critical habitat plays a major role in the protection plans for any endangered species, thus we hope our work can assist with future policy decisions surrounding the protection and management of the SRKW population.


\section{Building the modeling framework}

\subsection{Defining observer-animal encounters}

We define the movement trajectory through time of an individual animal $m$ as $\xi_m(t)$. This denotes the spatial coordinate of the individual at time $t$ with respect to the coordinate reference system used. We assume that there exists sufficiently large time windows $T_l \subset \mathbb{R} : l \in L$ such that the movement process driving the trajectories of the individuals $\xi_m(t) : t \in T_l$ have stationary invariant densities for each $T_l$, denoted $\pi_m(\textbf{s}, T_l)$. These densities define the UDs we aim to estimate and are assumed a-priori to be arbitrarily complex and represent the long-run density of locations at which the individual visits during $T_l$. An example time window could be a calendar month. 

Observers fall into two categories: static observers (e.g. hydrophones and camera traps) and mobile observers that may move through continuous space through time (e.g. vehicle-based observers). For each observer $o \in O$, we denote their position and field-of-view at time $t$ as $\xi_o(t)$ and $\phi_o(t)$ respectively. Unlike with $\xi_o(t)$, which defines a unique point in space at each time $t$, $\phi_o(t)$ defines a unique region or line. Thus $\phi_o(t) \subset \Omega \hspace{0.1cm} \forall t$ is a subset of the study region $\Omega$. The fields-of-view from two observers $o, \Tilde{o} \in O$ overlap in $T_l$ if $\phi_o(t) \bigcap \phi_{\Tilde{o}}(t) \neq \emptyset$ for some $t \in T_l$. We now describe the assumed data generating mechanism.

\begin{figure}[ht]
    \centering
    \includegraphics[scale=0.2]{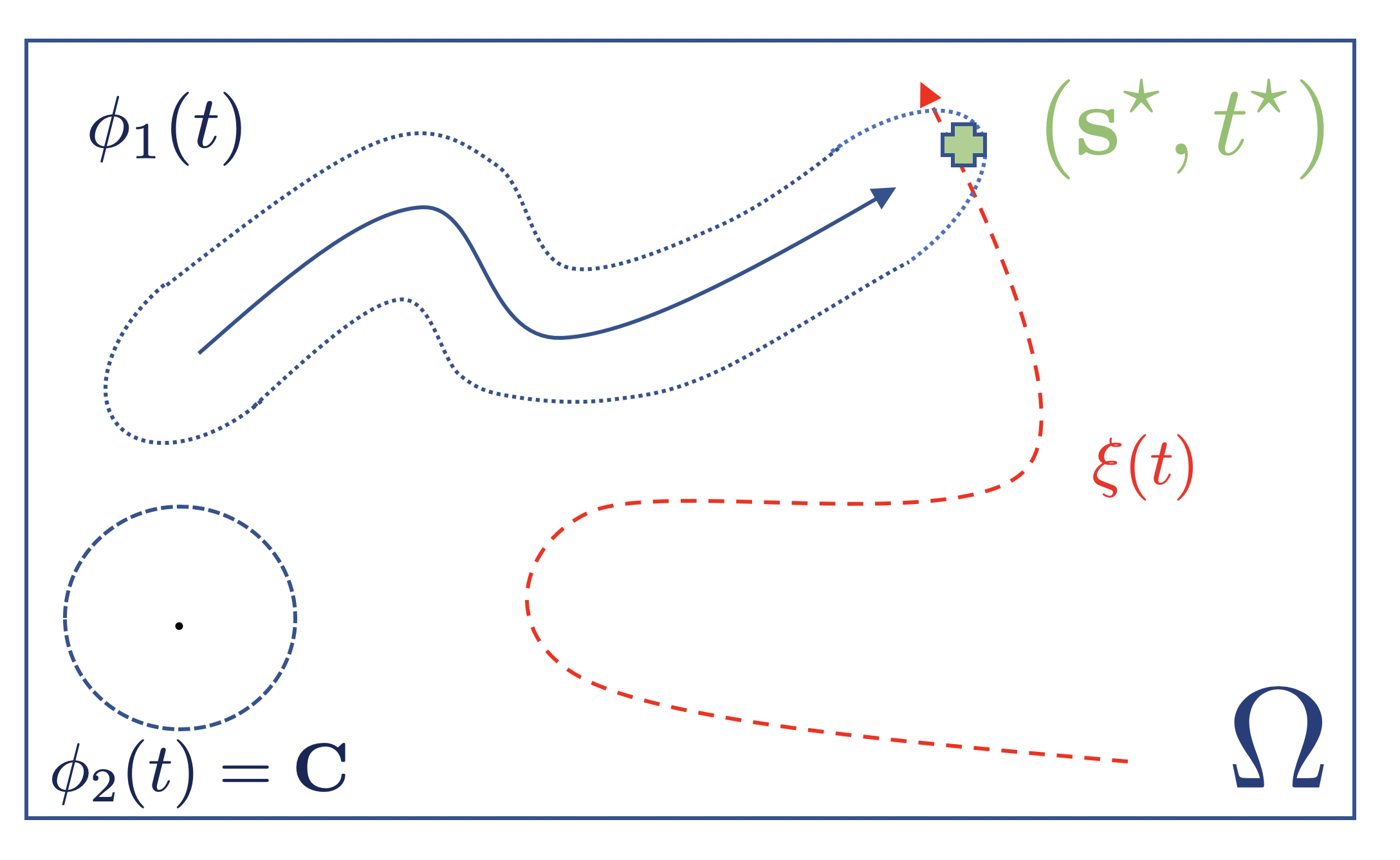}
    \caption{A diagram showing an example of an `encounter'. Two observers, one mobile and one static, with circular space-time fields-of-view $\phi_1(t)$ and $\phi_2(t)$ respectively are plotted with their fields-of-view through time shown as blue dotted lines. Both observers search for an individual moving with space-time trajectory $\xi(t)$ throughout the study region $\Omega$ shown as a red dashed line. The arrows denote the direction of travel. At time $t^\star$, the individual crosses into the mobile observers field-of-view at location $\textbf{s}^\star$. Thus, at time $t^\star$ the individual is encountered at location $\textbf{s}^\star$. Formally, at time $t^\star$, $\xi(t^\star) \in \phi_1(t^\star) \subset \Omega$. }
    \label{fig:encounter}
\end{figure}

Under the assumption of perfect detectability, and assuming the observers are searching continuously throughout $T_l$, we say that an encounter of individual $m$ occurs during time window $T_l$ when $m$'s movement trajectory $\xi_m(t)$ intersects with one or more observer's field-of-view function $\phi_o(t)$ for some time $t \in T_l$. That is $\xi_m(t) \in \phi_o(t)$ for some $t \in T_l$ and $o \in O$ during the study. When observers only search during a discrete set of times $t_j \in T_l : j \in \{1,...,J\}$, we say that a sighting occurs at time $t_j$ if $\xi_m(t_j) \in \phi_o(t_j)$. The assumption of perfect detectability by an observer within their field-of-view may be unrealistic and we allow for this assumption to be relaxed later. Fig \ref{fig:encounter} presents an example diagram displaying a setting with one moving individual, with two observers (one static, one mobile) searching for it. The individual intersects observer 1's field-of-view at time $t^\star$ and encountered at $\textbf{s}^\star$.



\subsection{Target point process model}

We assume that there exists an effort surface, denoted $\lambda_{eff}(\textbf{s}, T_l)$, that is able to fully capture the observers' fields-of-views and thus efforts. We assume that by conditioning on $\lambda_{eff}(\textbf{s}, T_l)$, the UD $\pi_m(\textbf{s}, T_l)$ can be recovered, with any observer bias removed. We use the following inhomogeneous Poisson process (IPP) model to describe the true data generating mechanism. We assume that the number of encounters of $m$ within any subregion $A \subset \Omega$ and time window $T_l \subset \cal T$ is Poisson-distributed with mean $\Lambda_{obs}(A, T_l) = \int_A \lambda_{obs}(\textbf{s},T_l) \textrm{d}\textbf{s} = \int_A \lambda_{true}(\textbf{s},T_l) \lambda_{eff}(\textbf{s},T_l) \textrm{d}\textbf{s}$ . We refer to $\lambda_{true}(\textbf{s},T_l)$ as individual $m$'s true intensity surface. With $\lambda_{eff}(\textbf{s}, T_l)$ able to fully capture the effort, $\pi_m(\textbf{s}, T_l) \propto \lambda_{true}(\textbf{s}, T_l)$. At location $\textbf{s}$ during time window $T_l$, it is defined as:
\begin{align*}
    \lambda_{true}(\textbf{s}, T_l) = \underset{\epsilon \to 0}{\lim} \hspace{0.1cm} E\left[N(B_\epsilon(\textbf{s}), T_l)\right] / \hspace{0.05cm} |B_\epsilon(\textbf{s})|,
\end{align*}
where $B_\epsilon(\textbf{s})$ denotes a circle with centre $\textbf{s}$, radius $\epsilon$, and area $|B_\epsilon(\textbf{s})|$. $N(B_\epsilon(\textbf{s}), T_l)$ denotes the number of encounters with $m$ per unit effort within the circle during time window $T_l$. Given the above assumptions, $\lambda_{true}(\textbf{s}, T_l)$ loosely represents the expected number of encounters with individual $m$'s trajectory within an infinitesimally small region around point $\textbf{s}$, per unit effort. A flat $\lambda_{true}(\textbf{s}, T_l)$ throughout $\Omega$ implies that $m$ exhibits complete spatial randomness throughout $\Omega$. Conversely, regions of high $\lambda_{true}(\textbf{s}, T_l)$ indicate `hotspots' of $m$.

Conditioned upon knowing $\lambda_{obs}(\textbf{s}, T_l)$ and observing a point pattern consisting of $n_{T_l}$ points (i.e. a collection of $n_{T_l}$ encounter locations) within the time set $T_l \in \cal T$, $\textbf{Y}_{T_l} = \{\textbf{s}_{i} : i \in \{1,...,n_{T_l}\}, \textbf{s}_{i} \in \Omega\}$, the likelihood of the spatio-temporal IPP is: 
\begin{align}
    \pi \left(\textbf{Y}_{T_l} | \lambda_{obs} \right) = \textrm{exp}\left\{ |\Omega| \int_\Omega \lambda_{obs}(\textbf{s},T_l)\textrm{d}\textbf{s}\right\} \prod_{\textbf{s}_i \in \textbf{Y}_{T_l}}\lambda_{obs}(\textbf{s}_i, T_l) \label{eq:LGCPlikelihood}
\end{align}
where $|\Omega|$ denotes the area of the domain $\Omega$.  


In practice, we will not know the effort intensity surface $\lambda_{eff}(\textbf{s}, T_l)$ and we will need to estimate it. When encounters are made in continuous time, and when overlap exists between the observers' fields-of-view, this quantity may be very complex. In this setting, accurately modeling this term would require the explicit modelling of the animal movement model within the observer's sampling process \citep{glennie2020incorporating}. In this paper, we show that even crude approximations of $\lambda_{eff}(\textbf{s}, T_l)$ can improve the statistical inference of $\pi_m(\textbf{s}, T_l)$. The crude approximation we use is based on the estimated path integrals of the observers' fields-of-view $\phi_o(t)$. When the locations of the observers are known (e.g. GPS positions), this can be computed by estimating $\phi_o(t)$ around each recorded location and then summing through time. When GPS positions are unavailable, the path integral can still prove a useful target quantity for building an effort emulator, or an effort model from a set of informative covariates. Let $|\cdot|$ denote the area or length function. Then, assuming no overlap exists between observers, the path integral approximation to $\lambda_{eff}(\textbf{s}, T_l)$ within a region $A_i \subset \Omega$ is $\int_{T_l} \sum_{o \in O} | \phi_o(t) \bigcap A_i | \textrm{d}t$. When the entirety of $A_i$ is observed throughout $T_l$ by an observer $o$, the estimated cumulative effort in $A_i$ from $o$ becomes $|A_i||T_l|$. The degree of approximation error will depend on many factors including the size of the fields-of-view relative to the size of $\Omega$, the number of observers, the accuracy in estimates of $\phi_o(t)$, and whether or not encounters were made in continuous-time or discrete-time. 

\subsection{Log-Gaussian Cox processes as a suitable base model for estimating UDs}

Linking an animal's space use to a set of covariates (e.g. sea surface temperature) is often a key component of an ecological analysis. The estimated relationships between encounter rate and environment allows researchers to predict variation in space use across space and time, possibly extrapolating into areas beyond the study area $\Omega$, and into time windows beyond the temporal domain $\cal T$. As with many popular regression-based SDM methods (linear models, GLMs, GAMs, etc.), $\lambda_{true}(\textbf{s}, T_l)$ may be modeled with a collection of nonlinear transformations of covariates, interactions, and splines within a log linear model: \begin{align}
   \textnormal{log } \lambda_{true}(\textbf{s}, T_l) = \boldsymbol{\beta}^T\textbf{x}(\textbf{s}, T_l), \label{eq:IPPlinearpred}
\end{align}
where $\textbf{x}(\textbf{s}, T_l)$ denote the set of measured covariates at location $\textbf{s} \in \Omega$ assumed constant throughout time window $T_l \subset \cal T$.
This IPP may be inadequate for use in ecological settings \citep{pacifici2017integrating}. The Poisson distribution assumed on the counts inside any subregion $A \subset \Omega$, implies the variance of the counts is equal to the mean. If the amount of environmental variability not captured by the modeled covariates is high, and the overdispersion is not controlled for, model-based confidence intervals can become overly-narrow and suffer from poor frequentist coverage \citep{baddeley2015spatial}. Spurious `significance' between the associations of covariates and the intensity may then be reported, unless computationally-intensive resampling methods, such as block-bootstrap, are performed \citep{fithian2015bias}.  

Cox process models extend the IPP by treating the intensity surface as a realisation of a random field \citep{baddeley2015spatial}. This enables the variance-mean relationships of the point process models to be more flexible. The random fields from the Cox process models can be specified to capture spatial, temporal, and/or spatio-temporal correlations, helping to control for any unmeasured covariates and biological processes driving the true intensities of the studied individuals \citep{yuan2017point}.



A popular class of flexible random fields chosen are Gaussian (Markov) random fields, letting $\lambda_{true}(\textbf{s}, T_l)$ be a realisation of a log-Gaussian process \citep{simpson2016going}. These models are called log-Gaussian Cox processes (LGCPs). Specification of a LGCP is achieved by adding a Gaussian process, denoted as $Z(\textbf{s}, T_l)$, to the linear predictor in (2):

\begin{align}
    \textrm{log} \hspace{0.1cm} \lambda_{true}(\textbf{s}, T_l) &= \boldsymbol{\beta}^T \textbf{x}(\textbf{s}, T_l) + Z(\textbf{s}, T_l), \label{eq:LGCPlinearpred}  \\
    \textbf{Z}(\textbf{S}, \textbf{T}) = \left[Z(\textbf{s}_1, T_{l(1)}),...,Z(\textbf{s}_n, T_{l(n)})\right]^T &\backsim N\left( \textbf{0}, \Sigma \right),\label{GRF}
\end{align}

where $\Sigma$ denotes the variance-covariance matrix of the Gaussian process evaluated at all of the $n = \sum_{l \in L} n_l$ locations and time windows $(\textbf{s}_i, T_{l(i)})$. The function $l(i)$ maps each observation to its corresponding time window. Different choices of covariance structures, lead to Gaussian processes with fundamentally different properties and uses. R packages such as spatstat and inlabru can fit such models \citep{baddeley2014package,inlabru}. We choose the LGCP as the base model for our framework due to its flexibility.  


\subsection{Covariates to account for detection probability and observer effort}

In practice, we can rarely satisfy the previous assumption of perfect detectability. Thus, we relax this assumption now and assume the existence of a `detection probability surface', $p_{det}(\textbf{s}, T_l)$ that can describe the heterogeneous detectability within the earlier point process model. Additional covariates may be available that can model both the heterogeneous detectability \citep[e.g. visibility indices, distance from the observer, etc.,][]{fithian2015bias} and/or the heterogeneous cumulative effort of each observer (e.g. distance from the nearest road). If these covariates are included in their correct functional forms, this regression adjustment approach may help to capture some of the heterogeneity in the observer effort and partially remove the associated biases \citep{dorazio2014accounting}. For applications with strongly informative covariates, such approaches have been shown to significantly improve predictive performance in SDMs \citep{elith2007predicting,fithian2015bias}. 

We model $p_{det}(\textbf{s}, T_l)$ with a set of covariates, $\textbf{w}_1(\textbf{s}, T_l)$, that are believed to affect only the observer abilities and not influence the true intensity of the individual of interest. These covariates are assumed constant throughout each time window $T_l$. By definition, the values of $p_{det}(\textbf{s}, t)$ are constrained to lie between 0 and 1. A value of 1 implies that all instances where the individual's trajectory intersects an observer's field-of-view leads to a recorded encounter. This could reflect a scenario where an easily-detected individual is in the immediate proximity to an observer under perfect weather conditions. A value less than 1 implies that some encounters are missed or not recorded. Thus, this helps to capture the processes that drive the under-reporting seen in many ecological studies. In the context of point processes, $p_{det}(\textbf{s}, t)$ is known as a thinning function. $T_l$-average sea state and $T_l$-average visibility are two example covariates that could be included in $\textbf{w}_1(\textbf{s}, T_l)$. Distance sampling functions can also be included following the approach of \citet{yuan2017point}. 

Similarly, we can model $\lambda_{eff}(\textbf{s}, T_l)$ with a set of covariates, $\textbf{w}_2(\textbf{s}, T_l)$, within a loglinear model. As before, the covariates used to explain observer effort are assumed to not directly impact the true intensity of the animals. This time, these covariates are believed to explain both the spatial distributions of observers and explain any differences in their efficiencies. When differences in observer efficiency exist, the earlier approximation to $\lambda_{eff}(\textbf{s}, T_l)$ becomes $\int_{T_l} \sum_{o \in O} \omega_o |\phi_o(t) \bigcap A_i | \textrm{d}t$, with the relative efficiencies captured in their weights $\omega_o$. This can help with the selection of relevant covariates and with model formulation. We advise modelling the observer efficiencies $\omega_o$ within the loglinear model and not the earlier probability surface to avoid an upper bound being placed on the efficiency of an observer. This allows for the existence of observers with higher skill levels than those who collected the data used to fit the model. Distance from road and observer type are two example covariates.  


Our joint model for true species' intensity, detection probability and observer effort is:  

\begin{align}
    \lambda_{obs}(\textbf{s}, T_l) &=\lambda_{true}(\textbf{s}, T_l)
    p_{det}(\textbf{s}, T_l)
    \lambda_{eff}(\textbf{s}, T_l) \label{eq:lambdaobs_noE} \\
    g^{-1} \left( p_{det} \right) (\textbf{s}, T_l) &= \boldsymbol{\gamma}_1^T \textbf{w}_1(\textbf{s}, T_l) \label{eq:lambdadet} \\
    \textrm{log } \lambda_{eff}(\textbf{s}, T_l) &= \boldsymbol{\gamma}_2^T \textbf{w}_2(\textbf{s}, T_l) \label{eq:lambdaeff} \\
    \textrm{log } \lambda_{true}(\textbf{s}, T_l) &= \boldsymbol{\beta}^T \textbf{x}(\textbf{s}, T_l) + Z(\textbf{s}, T_l),\label{eq:lambdatrue}
\end{align}


\noindent with $g$, a suitable link function (e.g.\ the logistic function), mapping the linear predictor of the detection probability surface to the unit interval. By assumption, $\pi(\textbf{s}, T_l) = \lambda_{true}(\textbf{s}, T_l)/\int_\Omega \lambda_{true}(\textbf{x}, T_l) d\textbf{x}$. 


Suppose the encounters are made by a collection of observers $o \in O$ with their unique observer efficiencies modeled with unique intercepts with respect to a baseline observer type. Then, the $\lambda_{true}(\textbf{s}, T_l)$ being modeled is interpreted as the expected encounter rate at location $\textbf{s} \in \Omega$ during $T_l$ of the individual by the chosen baseline observer. The log linear model then ensures that any differences in the observer efficiencies are modeled multiplicatively. Crucially, the interpretation of $\lambda_{true}(\textbf{s}_1, T_l) = 2 \lambda_{true}(\textbf{s}_2, T_l)$ is that the individual will occupy the area immediately around $\textbf{s}_1$ twice as often as around $\textbf{s}_2$ in the long run. Finally, if a series of encounters/detections are made where the entirety of $\Omega$ is perfectly observable, then both $p_{det}(\textbf{s}, T_l)$ and $\lambda_{eff}(\textbf{s}, T_l)$ should be fixed equal to a constant. Two examples are when an ultra high-resolution satellite image containing the entirety of $\Omega$ is taken and where telemetry data is available. In the latter case, care must be taken to properly subset the telemetry data to ensure that any autocorrelations from the individual's movement removed.

Estimation of the non-intercept terms within the detection $\boldsymbol{\gamma}_1$, effort $\boldsymbol{\gamma}_2$ and the environmental $\boldsymbol{\beta}$ parameters is possible, so long as all the corresponding covariates $\textbf{x}$, $\textbf{w}_1$, and $\textbf{w}_2$ are not linearly dependent or interact \citep{dorazio2014accounting}. Non-perfect correlation between the three sets of covariates, whilst making estimation more difficult, does not affect the identifiability of these parameters \citep{fithian2015bias}. This is a desirable property in the context of UDs. The intercept parameters are estimable if either there is independence between $\textbf{x}$, $\textbf{w}_1$ and $\textbf{w}_2$, or if at least one accurate control dataset is included in the joint model \citep{fithian2015bias}. A control could be a survey with known observer effort. When the intercept is desired, the animal movement process should be considered to reduce bias \citep{glennie2020incorporating}.  Furthermore, the estimability of observer-specific intercepts (i.e. relative observer efficiencies) will require significant spatial-overlap to exist between the cumulative observer efforts of the different observers. This is especially true when spatially correlated $Z(\textbf{s}, T_l)$ terms are included, since any differences in observer efficiencies may be erroneously captured by the $Z(\textbf{s}, T_l)$ terms. Of course, the assumption of non-overlapping fields-of-view at all times $t \in T_l$ is still required. If the above conditions hold, then after model-fitting, fixing both sets of covariates $\textbf{w}_1(\textbf{s}, T_l), \textbf{w}_2(\textbf{s}, T_l)$ equal to a constant allows the effects of variable detection probability and observer effort to be removed from predictions throughout $\Omega$.

\subsection{Approximating effort from GPS-tagged observers}

In many situations, $\int_{T_l} |\phi_o(t) \bigcap A_i| \textrm{d}t$ may be known or directly estimable for a set of pixels $A_i \subset \Omega$ used to approximate (1). For example, mobile observers may record their GPS coordinates, or may be known to travel along a strict network of known routes (e.g. shipping lanes). Similarly, the locations of static observers (e.g.\ camera traps) and their detection ranges may also be known. In both cases, a simple, yet effective approach for approximating $\lambda_{eff}(\textbf{s}, T_l)$ can be implemented. Given $\phi_o(t)$, we can include $|A_i|^{-1} \int_{T_l} |\phi_o(t) \bigcap A_i| \textrm{d}t$ as a fixed covariate within $\textbf{w}_2$ to represent the average effort within $A_i$. Then, only the corresponding slope term $\omega_o$ in $\boldsymbol{\gamma}_2^T$ needs to be estimated. When only one observer type is available, including the logarithm of $|A_i|^{-1} \int_{T_l} |\phi(t) \bigcap A_i| \textrm{d}t$ as an offset (i.e.\ fixing $\omega \equiv 1$) is all that is required.  


The performance of the above approach may deteriorate as the degree of overlap between the fields-of-view of the observers increases. One solution is to remove the effort and encounters from overlapping observers. Alternatively, it may be possible to model the inter-observer autocorrelations directly \citep{clare2017pairing}. We demonstrate the utility of the path integral approximation in a simulation study in Section 4.

\subsection{Generalising the base model with the addition of marks}

Often, the utilization distributions of multiple individuals of a species or population and their changes through time are desired \citep{elith2007predicting, fithian2015bias}. Furthermore, understanding the factors driving the individuals to use the space may also be of importance to researchers. Spatio-temporal point processes can be further generalized to marked spatio-temporal point processes to allow for a greater range of research questions to be tackled \citep[e.g.][]{chakraborty2011point}. The main idea of marked point processes is that for each point, we observe attributes in addition to its location and time. These attributes are called marks. Marks might be categorical variables such as whale pod or an indicator of foraging behaviour, count variables such as group size, or continuous variables such as travel speed.  


Formally, we associate a random variable $m_y$ (a mark) to each location and time window of the random set $\textbf{y} \in Y_{T_l}$. We place a probability distribution on each of the marks, and model the joint distribution of the locations and marks. Let $M$ denote the support of a distribution of marks $m_y$. The mark distribution is allowed to depend upon space and time (i.e. depend upon $\textbf{y}$ and $T_l$), but is not allowed to depend on other points in $Y_{T_l}$. Thus, the $m_y$ for different $\textbf{y} \in Y_{T_l}$ are independent. Now the pair $\left(Y_{T_l}, m_Y \right)$ may be viewed as a random variable $Y_{T_l}^\star$ in the product space $\Omega \times M$. There is no limit to the number of marks that can be associated with each point. We simply need to include a probability distribution for each of the $J$ marks $m_{Y_{T_l}}^j : j \in \{1,...,J\}$. 

When the mark distribution of one of the $J$ marks is discrete (i.e. when $m_{Y_{T_l}}^j \in \{1,...,K\}$), as in the case of individual ID, we can estimate the probability that the presence of an individual at a given location $\textbf{s} \in \Omega$ within $T_l \subset \cal T$ has $j^{th}$ mark equal to $k \in \{1,...,K\}$. For notational simplicity, let $J = 1$ and define $\lambda_{true} (\textbf{s}, T_l, k)$ to be the true intensity for the mark category $k$ during $T_l$ (i.e. $\left\{ \lambda_{true}(\textbf{s}, T_l, m) : m_{\textbf{s},T_l} = k \right\}$). The probability at location $\textbf{s}$ is then:

\begin{align}
    p_{true}(\textbf{s}, T_l, k) &= \frac{ \lambda_{true}(\textbf{s}, T_l, k) }{ \sum_{\kappa=1}^{K} \lambda_{true}(\textbf{s}, T_l, \kappa) } \nonumber \\ 
    &= \frac{\textrm{exp} \left( Z(\textbf{s}, T_l, k) + \boldsymbol{\beta}^T \textbf{x}(\textbf{s}, T_l, k) \right)}{\sum_{\kappa=1}^{K} \textrm{exp} \left( Z(\textbf{s}, T_l, \kappa) + \boldsymbol{\beta}^T \textbf{x}(\textbf{s}, T_l, \kappa) \right) } \label{eq:probobskequation}
\end{align}




In many cases, computing and plotting estimates of $p_{true}(\textbf{s}, T_l, k)$, the true mark-specific probabilities will be the inferential target. When $k$ denotes the individual ID, estimates of $p_{true}(\textbf{s}, T_l, k)$ can help to establish spatial niches specific to the $k^{th}$ individual, whilst removing any observer and detectability biases. Note that when $K$ contains all individuals of a target population and when $\Omega$ is sufficiently large, then the normalized denominator of (\ref{eq:probobskequation}) reflects the distribution of the whole population. When some or all of the parameters and/or covariates are shared between the mark-specific intensities,  cancellations will occur in (\ref{eq:probobskequation}). 

\subsection{The proposed model framework}

Let $\Omega$, $\cal T$, and $T_l \subset {\cal{T}} : l \in L$ be defined as before. Let the support of the marks be $M$. Suppose for each $T_l$ we have a collection of encounter locations and marks $(\textbf{Y}_{T_l}, \textbf{M}_{Y_{T_l}}) = \{ (\textbf{s}_i, \textbf{m}_i) : (\textbf{s}_i, \textbf{m}_i) \in \Omega \times M \}$. The proposed model is:

\begin{align}
    \lambda_{obs}(\textbf{s}, T_l, \textbf{m}) &=
   \lambda_{true}(\textbf{s}, T_l, \textbf{m})
    p_{det}(\textbf{s}, T_l, \textbf{m})
    \lambda_{eff}(\textbf{s}, T_l, \textbf{m})
   \label{eq:lambdaobsfullmodel}
    \\
    g^{-1} \left( p_{det}(\textbf{s}, T_l, \textbf{m}) \right) &= \boldsymbol{\gamma}_1^T \textbf{w}_1(\textbf{s}, T_l, \textbf{m}) \nonumber \\
    \textrm{log } \lambda_{eff}(\textbf{s}, T_l, \textbf{m}) &= \boldsymbol{\gamma}_2^T \textbf{w}_2(\textbf{s}, T_l, \textbf{m}) \nonumber \\
    \textrm{log } \lambda_{true}(\textbf{s}, T_l, \textbf{m}) &= \boldsymbol{\beta}^T \textbf{x}(\textbf{s}, T_l, \textbf{m}) + Z(\textbf{s}, T_l, \textbf{m}). \nonumber
\end{align}

\begin{figure}[ht!]
    \centering
    \includegraphics[trim={0.07cm 1cm 0 0},clip,scale=0.3]{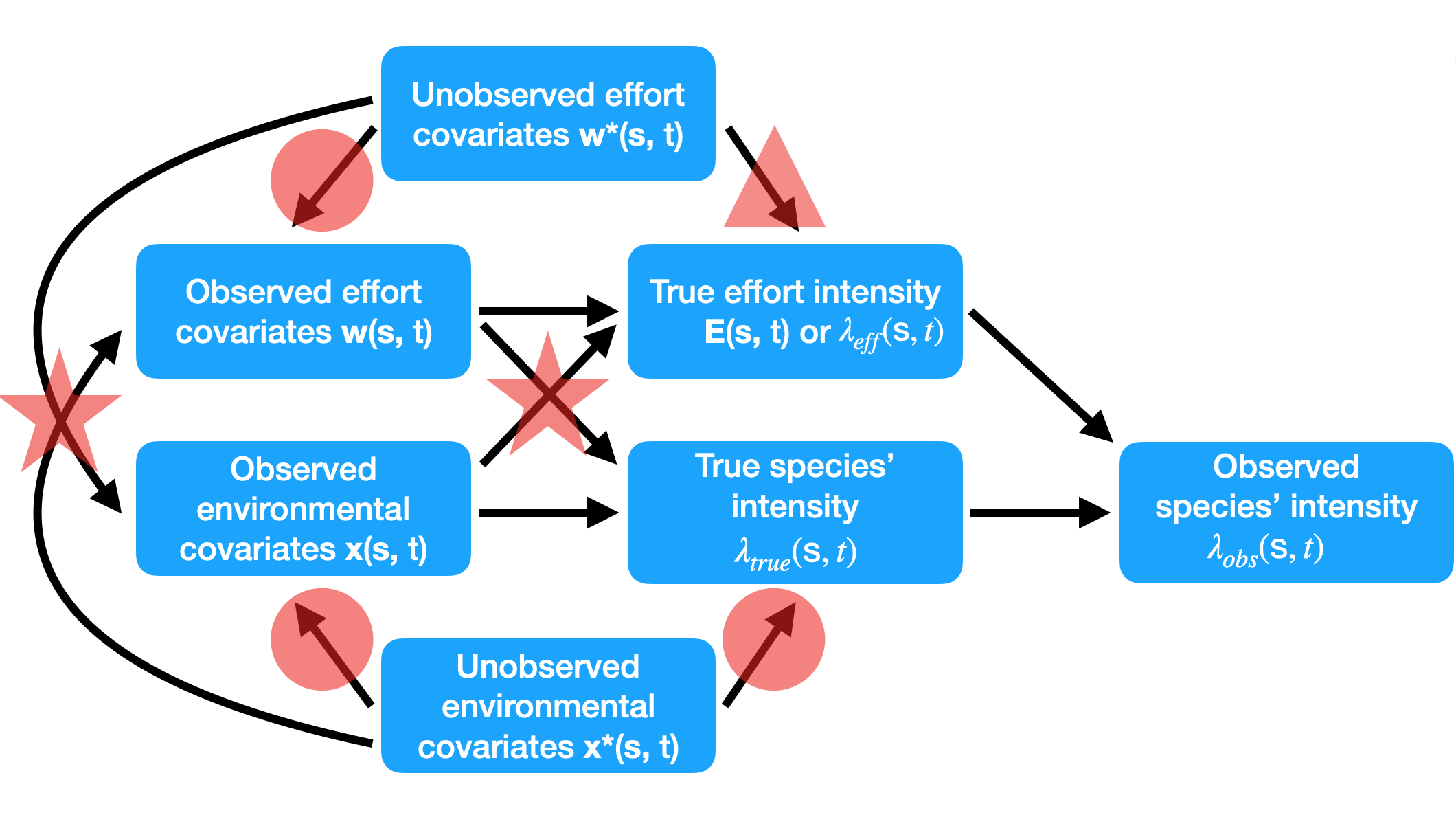}
    \caption{A plot showing the assumed causal DAG for the proposed framework with the detection probability assumed constant. An arrow between a variable set $A$ and a variable set $B$ indicates that at least one variable exists in both sets with a direct causal effect between them. The causal Markov assumption is made such that a variable is independent of its non-descendants, when conditioned on its parents \citep{hernan2010causal}. If any of the causal effects found within the red shapes exist, then problematic confounding may follow. This is explained at depth in the supplementary material. }
    \label{fig:causal_dag}
\end{figure}

For estimates of $\lambda_{true}(\textbf{s}, T_l, \textbf{m})$ under the above framework to be free from confounding by effort and for estimates of individual-environment relationships to be accurate, many assumptions are required in addition to those highlighted earlier. As shown in the causal directed acyclic graph (DAG) in Fig \ref{fig:causal_dag} \citep{hernan2010causal}, one of the fundamental assumptions required for estimates of $\lambda_{true}(\textbf{s}, T_l)$ to be free of confounding, is that $\lambda_{eff}(\textbf{s}, T_l, \textbf{m})$ fully describes the efforts of the observers through space and time across the marks. This is achieved when either the covariates $\textbf{w}_2(\textbf{s}, T_l, \textbf{m})$ completely explain the efforts of the observers, or when known or accurate estimates of observer effort are included in $\textbf{w}_2(\textbf{s}, T_l, \textbf{m})$. For estimates of the UD to be free of confounding, only the relative efforts of the observers need be known or estimable. In either case, with unknown effort, the existence of unobserved covariates of effort can confound estimates of $\lambda_{true}(\textbf{s}, T_l, \textbf{m})$ in two ways. 

Firstly, if the unobserved effort covariates affect one or more observed environmental covariates $\textbf{x}(\textbf{s}, T_l, \textbf{m})$, then the corresponding effect estimate $\boldsymbol{\beta}$ and hence the individual's intensity $\lambda_{true}(\textbf{s}, T_l, \textbf{m})$ may remain confounded by the effort. For example, suppose effort is unknown for a dataset containing encounters with marine-based individuals. Suppose that distance-to-shore is not included as a covariate despite strongly impacting search effort. Furthermore, suppose chlorophyll-A, which has high values closer to shore, is included in $\textbf{x}(\textbf{s}, T_l, \textbf{m})$. Estimates of the effects of chlorophyll-A in $\boldsymbol{\beta}^T$ will likely be confounded by effort, which in turn will bias the estimates of $\lambda_{true}(\textbf{s}, T_l, \textbf{m})$. 

Secondly, even if the unobserved effort covariates are independent of $\textbf{x}(\textbf{s}, T_l, \textbf{m})$, residual spatio-temporal correlations in the sightings data driven by the unobserved effort covariates may be erroneously captured by the Gaussian process $Z(\textbf{s}, T_l, \textbf{m})$. Consequently, estimates of $\lambda_{true}(\textbf{s}, T_l, \textbf{m})$ may therefore remain confounded by the heterogeneity in the observer effort. At this point, one may be tempted to simply include a unique Gaussian process for $\lambda_{eff}$ to capture these missed covariates. This cannot be done. Without additional knowledge available that can adequately constrain the additional Gaussian process, it will be non-identifiable. Similar problems occur if a detection probability surface $p_{det}(\textbf{s}, T_l, \textbf{m})$ is estimated. Once again, the true detectability of the species must be fully captured by $\textbf{w}_1(\textbf{s}, T_l, \textbf{m})$. 

Removing the confounding of $\lambda_{true}(\textbf{s}, T_l, \textbf{m})$ by effort will be challenging for many ecological applications and it highlights the need for increased collection of effort information along with sightings data. However, Fig \ref{fig:causal_dag} shows that if effort is known, then conditioning on it can remove all the problematic confounding at location $\textbf{s}$. Furthermore, in the simulation study in Section 4, we show that even crude estimates of observer effort and detectability can dramatically improve estimates of $\lambda_{true}(\textbf{s}, T_l, \textbf{m})$ compared with simply ignoring effort altogether. The issues of confounding are not exclusive to our framework and are present across all methods for estimating both UDs and SDMs \citep{fithian2015bias, koshkina2017integrated}. Thus, these concerns should not be seen as a weakness of our framework, but instead as a weakness inherent to biased data collection-protocols. 

If observer effort is limited to a small subregion $\Omega_0 \subset \Omega$, such that $\lambda_{eff}(\textbf{s}, T_l, \textbf{m}) = 0 \hspace{0.1cm} \forall \textbf{s} \in \Omega \cap \Omega_0^C$, then additional assumptions must be placed on how $\Omega_0$ was selected. For example, if search effort was focused in regions where the species' intensity was expected to be highest, then extrapolated estimates of $\lambda_{true}(\textbf{s}, T_l, \textbf{m})$ into the regions of zero effort $\Omega \cap \Omega_0^C$ may remain biased. Estimates of the true intensity $\lambda_{true}(\textbf{s}, T_l, \textbf{m})$ into these regions may be too high, with estimates of the intercept positively biased \citep{watson2018general}. This issue is known as preferential sampling \citep{pennino2019accounting}. This highlights the benefits of conducting high-quality surveys with randomised effort. By choosing $\Omega_0$ at random, no systematic bias is expected in predictions into $\Omega \cap \Omega_0^C$, and extrapolation of the intensity throughout $\Omega$ can be performed with greater confidence. When there is little-to-no confidence that $\Omega_0$ was selected in a random manner, predictions should be constrained to lie within $\Omega_0$.

An important advantage of the framework is that data from different observers and of differing type can be combined to jointly estimate one intensity surface \citep{koshkina2017integrated}. This is achievable as the intensity given by (\ref{eq:lambdaobsfullmodel}), can be linked to the likelihoods of several common data types, including aggregated forms. For example, the logistic regression likelihood for binary site presence-absence data, the Poisson likelihood for site count data and the LGCP likelihood for presence-only data can all be derived from the intensity (see \citet{hefley2016hierarchical} and the supplementary material). Distance sampling methods have also been fit using LGCPs (see \citet{yuan2017point} for details). All that is required for the suitable combination is that encounters with individuals are approximately independent snapshots of their UDs and that any heterogeneous effort or detectability can be suitably controlled for. This is a clear demonstration of the unifying potential of the LGCP for ecological data, and the causal DAG of Fig \ref{fig:causal_dag} is a useful tool to assess whether the assumptions needed to apply model (10) are satisfied. A short summary of how to approximate the LGCP likelihood and additional details on the DAG are provided in the supplementary material.

\section{Simulation study}

We now present a simulation study to demonstrate the ability of the framework to combine encounter data from mobile and static observers and predict an animal's UD with both minimal bias and high precision. We first simulate the movements of observers and an animal and generate encounters following the earlier data generating mechanism. Next, we use the recorded locations of the observers to compute the approximation to effort that was introduced earlier. We ignore issues of overlap. We then plug this estimate of effort in to a point process model. Note that we do not attempt to explicitly model the movements of the animal within the sampling process model as in \citet{glennie2020incorporating}. Despite the use of a crude approximation for effort, we show that the framework offers improvements in prediction performance, even when the analyst incorrectly specifies $\phi_o(t)$ throughout time. Based on these results, we provide a list of recommendations for analysts. 

\begin{figure}[ht!]
    \centering
    \includegraphics[scale=0.5]{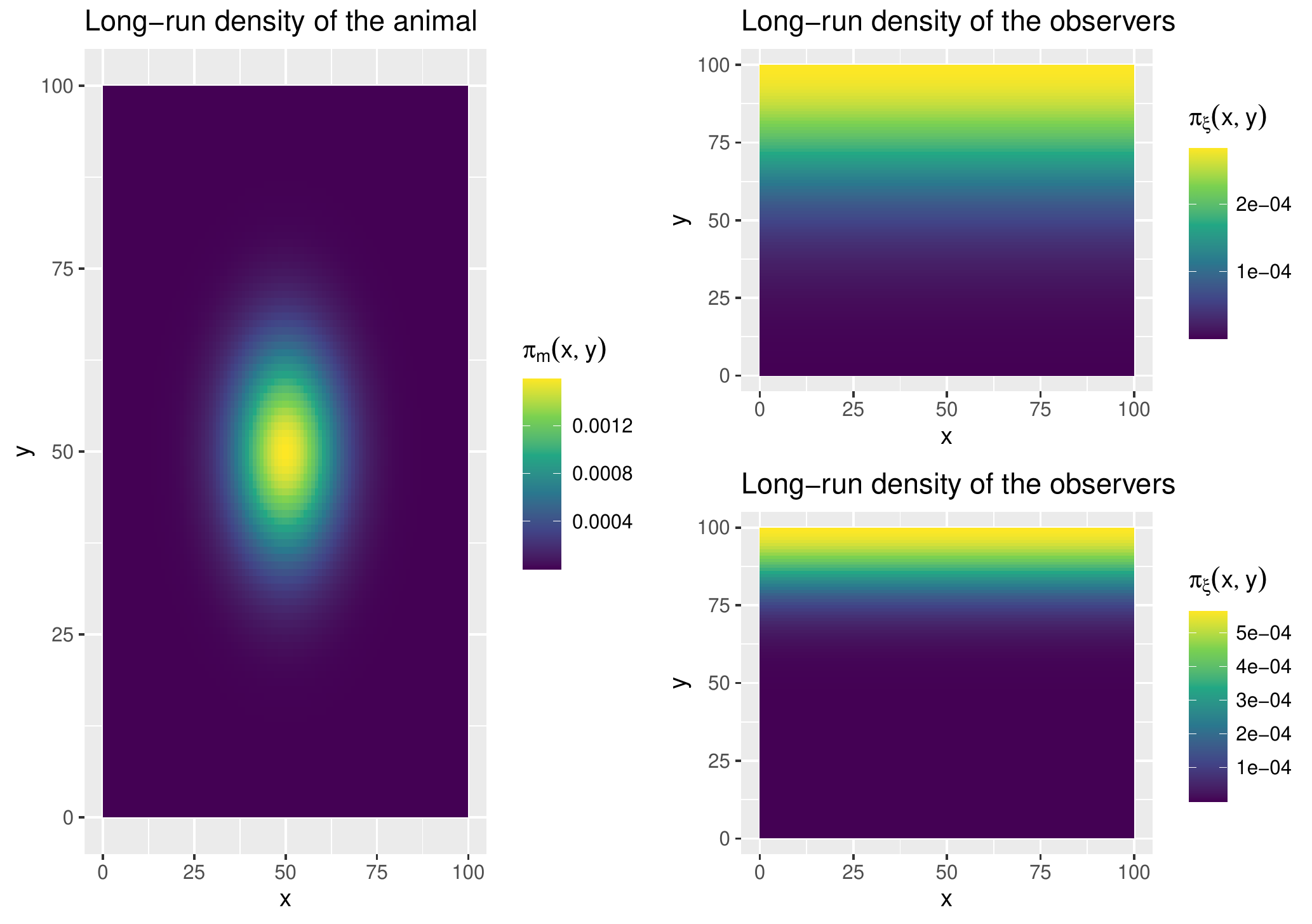}
    \caption{A plot showing the long run densities of the animal and the observers. The top-right and bottom-right plots show the low and high observer bias settings respectively. The smaller Brownian motion variance leads to a higher concentration of effort in the North of the study region and hence a larger degree of observer bias.}
    \label{fig:longrunsim}
\end{figure}

We simulate the movements of an animal and a set of observers using the model of \citet{brillinger2012use}. In particular, we use a stochastic differential equation (SDE) with potential functions chosen to ensure a desired long-run behaviour. The animal's potential function is chosen to be the logarithm of a symmetric bivariate normal distribution centered at $(\mu_x, \mu_y) = (50, 50)$, with variance 100. The observers' potential function is specified as the logarithm of a univariate half-normal distribution centered at $m_y=100$, with variance 200. The variance of the Brownian motion terms driving the movements is fixed at 2 for the animal, and fixed at either 2 or 8 for the observers. Thus, the animal's UD is a symmetric bivariate normal distribution, with the observers' UD a univariate normal distribution which focuses their efforts in the North of the study region (Fig \ref{fig:longrunsim}). The study region is a square with side lengths equal to 100 arbitrary units. The observers are given circular fields-of-view with maximum range of 10 units. These settings imply that the study region is very small.


We discretize time and use a first-order approximation to generate paths from the continuous-time SDE. Thus, both the simulated movements and potential encounter events occur across the discrete time-steps. The average distance travelled at each time step is roughly 1.75 units for the animal, and either 1.75 or 3.5 units for the mobile observers depending on whether the variance of the Brownian motion is 2 or 8. At each time step, if the animal is closer than 10 units of distance from an observer, it is encountered with a probability that decays linearly from 1 to 0 as the distance from the observer increases from 0 units to 10 units. If the animal is detected within 500 time-steps, representing a single `trip', then the encounter location is recorded along with the observers' tracks. If no encounter occurs during the trip, then only the observers' tracks are recorded. For each trip, we randomly sample the initial locations of the animal and the observers from their respective UDs. Subsequent locations are restricted from leaving the study region. For static observers, we simply hold their initial values fixed through time. The fields-of-view of all observers may overlap. 

For each simulation iteration, we then repeat the above steps 150 or 300 times to generate 150 or 300 trips. We fit a (IPP) point process model with correctly specified parametric form to the encounter locations with and without effort adjustment. We compute a crude approximation of effort. The observers' paths are mapped to a coarse $100 \times 100$ grid of pixels. Next, estimates of their fields-of-view, $\hat{\phi}_{o}(t)$, are computed, and then path integrals of $\hat{\phi}_{o}(t)$ are taken over the grid. Log-values of these path integral approximations are then included as an offset within the IPP. Here, estimates of effort are summed across the observers, ignoring any overlap in their fields-of-view. We also adjust for overlap in the supplementary material, but only minor improvements in predictive performance are seen. For model comparison, we compute at each simulation iteration both the mean squared prediction error (MSPE) of the animal's UD across the grid of pixels, and the bias of the estimated y-axis center of the UDs $\hat{\mu}_y$. The MSPE is computed with respect to the true UD. 

To understand how the method performs in practice, we change both the data-generating mechanism (DGM) and the assumptions made by the analyst when formulating estimates of observer effort. For the DGM, we adjust the degree of observer bias from `high' to `low' by changing the variance of the Brownian motion driving the observers' motions from 2 to 8 respectively. We also change the number and type of observers (mobile and/or static) and the number of trips made (150 or 300). For estimating effort, we either assume perfect detectability across $\hat{\phi}_{o}(t)$, or we model the linearly decaying distance sampling function. Next, we either underestimate, correctly specify, or overestimate the detection range of $\hat{\phi}_{o}(t)$ at 2, 10, and 50 units respectively. We perform 100 replications of each setting. 

\begin{figure}
    \centering
    \includegraphics[scale=0.7]{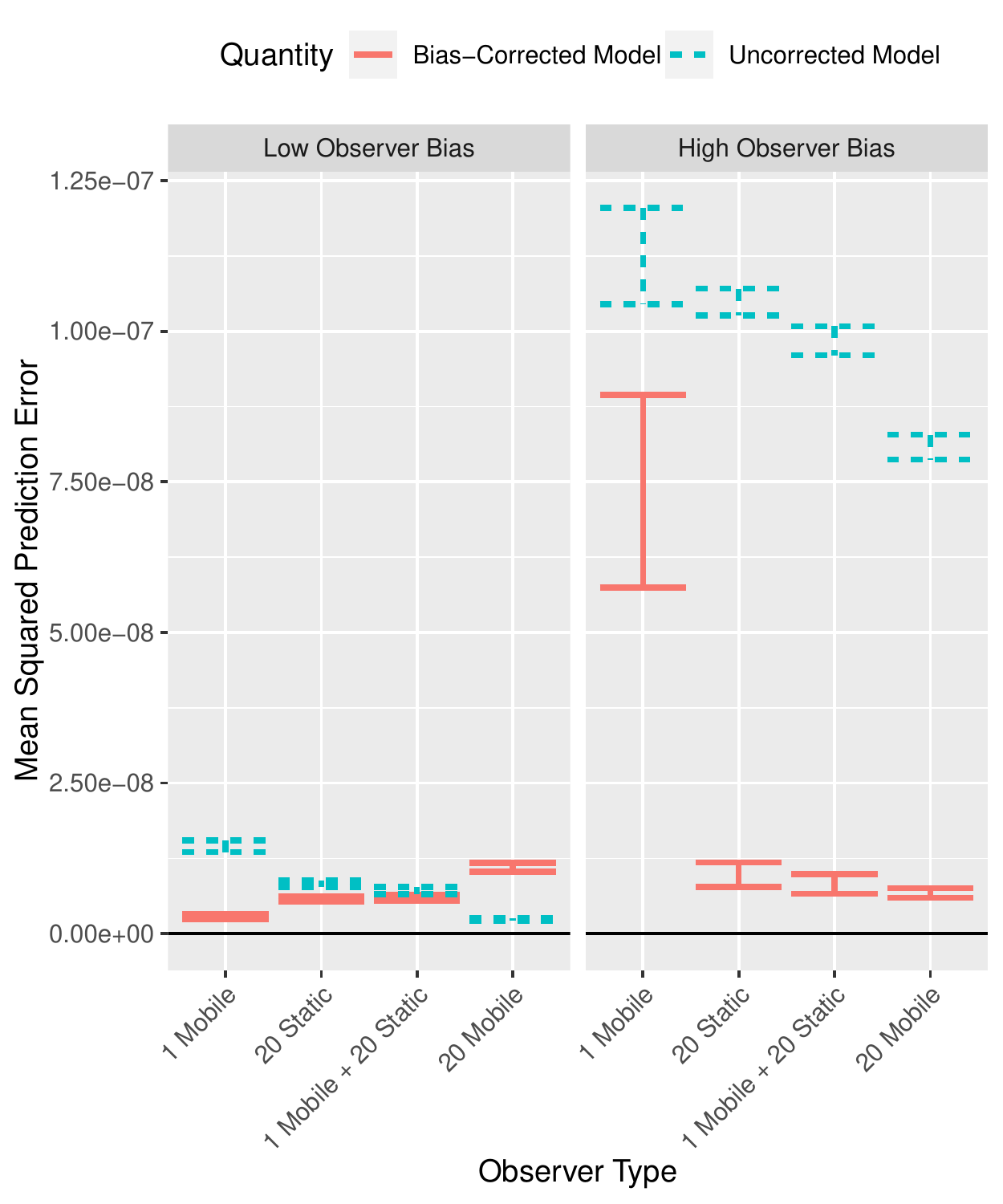}
    \caption{A plot showing the mean squared prediction error (MSPE) of the animals UD under the bias-corrected and bias-uncorrected models vs the types of observers. From left to right are the results from one mobile observer, twenty static observers, and twenty static with one mobile observers. The degree of observer bias is changed from low to high in the columns. The red solid lines and the blue dashed lines show the median MSPE along with robust intervals computed as $\pm 2c \textnormal{MAD}$ from the Bias-corrected and uncorrected models across the 100 simulation replicates respectively. The median absolute deviations about the medians (MAD) have been scaled by $c = 1.48$. This ensures that the intervals are asymptotically equivalent to the 95\% confidence intervals that would be computed if the MSPE values were normally distributed. Note that here all the analyst's assumptions correctly match the true data-generating mechanism, albeit with any overlap in the observers' efforts ignored.}
    \label{fig:APV_DGM} 
\end{figure}

\subsection{Effects of observer effort and detection range misspecification} Fig \ref{fig:APV_DGM} and Fig S\ref{fig:Bias_DGM} in the supplementary material demonstrate that improvements in both prediction variance and bias can be attained with the approximate effort-correction approach. These benefits are seen across all the observer types (i.e. static, mobile, and combinations) and the typical performance of the bias-corrected method is seemingly insensitive to the degree of the observer bias. In contrast, the performance of the uncorrected model is greatly affected by both the level of observer bias and observer type. In particular, the uncorrected model performs poorly when one ignores large observer bias. The results from both the 150 and 300 trip settings are similar and so we aggregate the results when forming the plots. 

\begin{figure}
    \centering
    \includegraphics[scale=0.7]{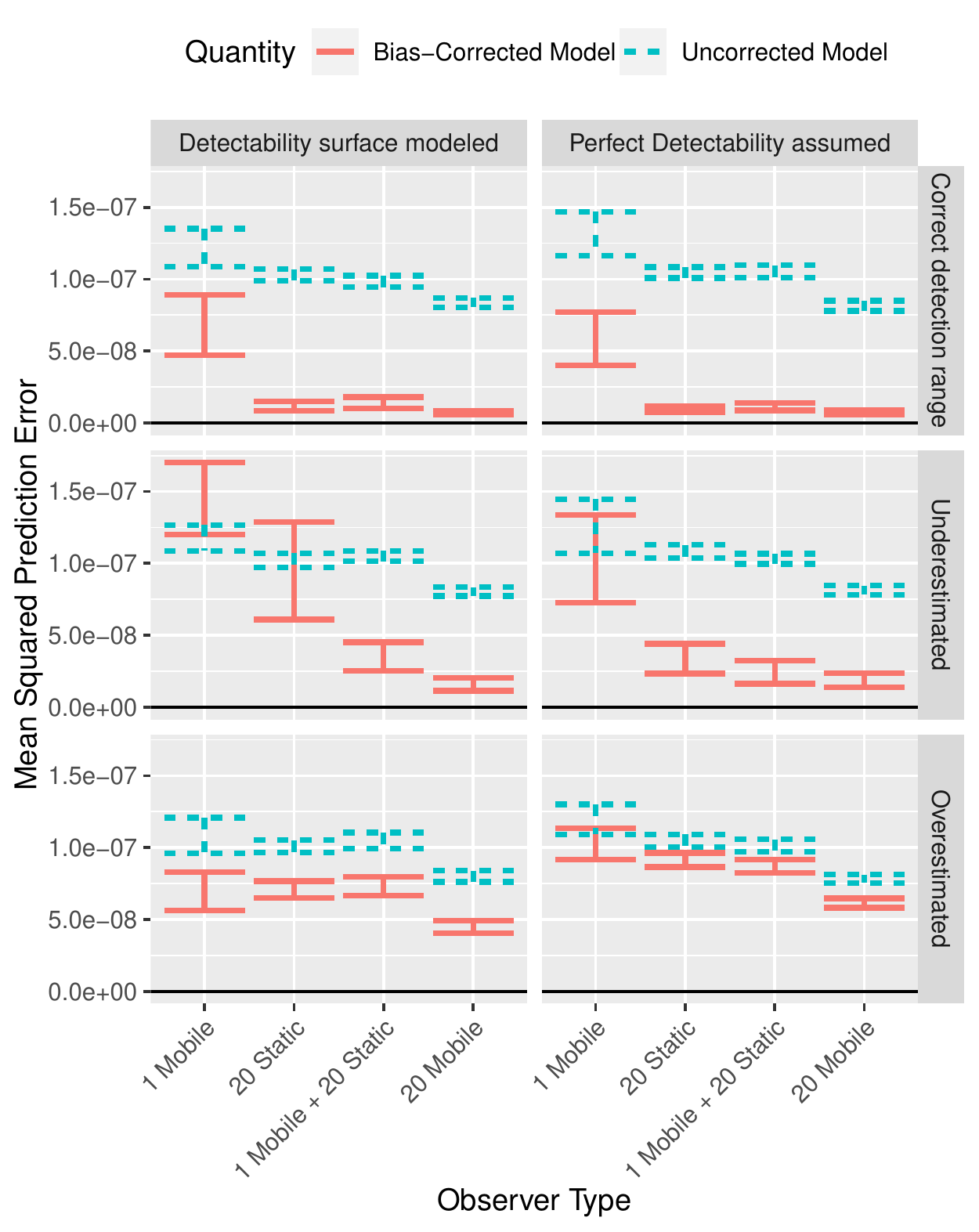}
    \caption{A plot showing the mean squared prediction error (MSPE) of the estimated animal's UD under the bias-corrected and bias-uncorrected models vs the types of observers. From left to right are the results from one mobile observer, twenty static observers, twenty static with one mobile observers, and twenty mobile observers. The distance sampling function has either been modeled or ignored in the two columns from left to right and the observers' detection range has been assumed to be 10, 2 and 50 across the rows. The red solid lines and the blue dashed lines show the median MSPE along with robust intervals computed as $\pm 2c \textnormal{MAD}$ from the Bias-corrected and uncorrected models across the 100 simulation replicates respectively. The MAD has been scaled by $c = 1.48$. This ensures that the intervals are asymptotically equivalent to the 95\% confidence intervals that would be computed if the MSPE values were normally distributed. Note that the results are shown for 150 trips with high observer bias.}
    \label{fig:APV_analyst}
\end{figure}

\begin{figure}
    \centering
    \includegraphics[scale=0.7]{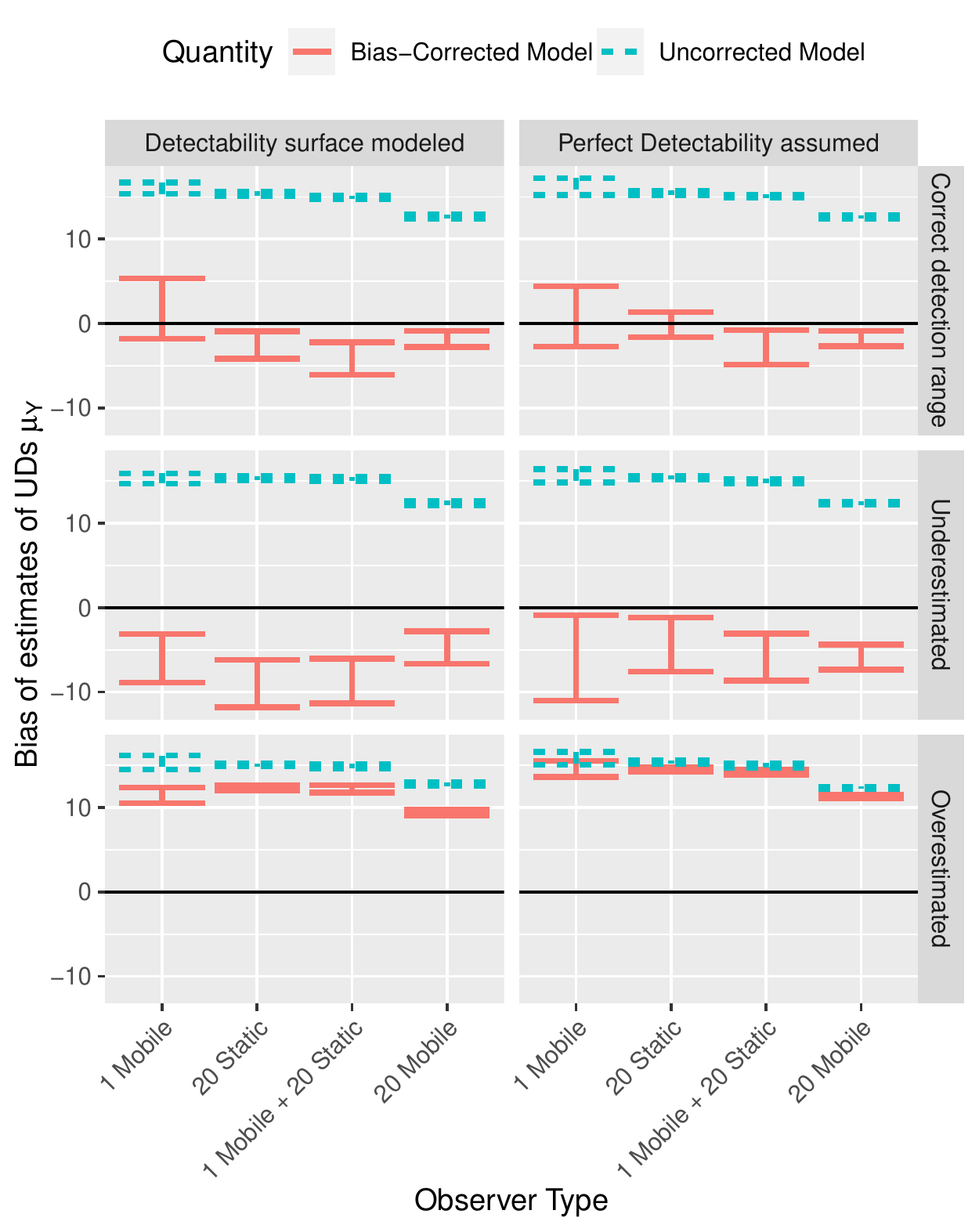}
    \caption{A plot showing the bias of the estimated animal's UD center $\mu_y$ under the bias-corrected and bias-uncorrected models vs the types of observers. From left to right are the results from one mobile observer, twenty static observers, twenty static with one mobile observers, and twenty mobile observers. The distance sampling function has either been modeled or ignored in the two columns from left to right and the observers' detection range has been assumed to be 10, 2 and 50 across the rows. The red solid lines and the blue dashed lines show the median bias along with robust intervals computed as $\pm 2c \textnormal{MAD}$ from the Bias-corrected and uncorrected models across the 100 simulation replicates respectively. The MAD has been scaled by $c = 1.48$. This ensures that the intervals are asymptotically equivalent to the 95\% confidence intervals that would be computed if the biases were normally distributed. Note that the results are shown for 150 trips with high observer bias. }
    \label{fig:Bias_analyst}
\end{figure}

The performance of the bias-correction approach is sensitive to the analyst using the correct detection ranges of the observers to define the observers' fields-of-view $\phi_o(t)$ (Figs S\ref{fig:APV_analyst} and S\ref{fig:Bias_analyst}). However, improvements in both the prediction variance and the bias of UD center-estimates can still be seen, even with badly misspecified detection ranges. Underestimating the observers' detection range leads to an over-correction of observer bias and leads to estimates of the animal's UD center to be negatively biased. The converse is true when the observers' ranges are overestimated. Both lead to increases in MSPE. Interestingly, the bias-correction method appears insensitive to whether or not a distance sampling function is used. 


The MSPE is a measure of predictive performance that is driven by both the squared prediction biases and the variances of the predictions. For the uncorrected model, the heterogeneous observer effort is the major cause of prediction bias. This is expected to decrease as the number of observers increases, due to the study region becoming increasingly explored. For the effort-corrected model, two major causes of prediction bias remain in our crude approach for approximating effort. The first such cause is due to the approximation error of using the path integrals of the fields-of-view to represent the cumulative effort. Even if $\phi_o(t)$ is correctly specified at all times, approximation error will remain due to the statistical dependence between the encounter/non-encounter events through time. To understand this, suppose an observer has failed to record an encounter for a significant period of time. Conditional upon this information, the current location of the animal is unlikely to be situated within the immediate proximity of the observer. Accurate estimates of the true effort $\lambda_{eff}(\textbf{s}, t)$ would need to be adjusted to account for this fact. This would require explicitly modelling the animal's movement process jointly within the sampling model \citep{glennie2020incorporating}. The second cause of prediction bias is due to overlap in the observers' fields-of-view. This gets worse as the density of observers increases. Multiple factors impact the variance of the predictions. For both models, the variance of the predictions decreases as the number of encounters increases. For the effort-corrected model, longer observer paths further reduce the variance. Both encounter frequency and the cumulative observer path length increase with the number of observers.

For the uncorrected model, we indeed see that the mean squared prediction error (MSPE) decreases as the number of observers is increased. The largest improvements are seen with the addition of mobile observers due to the study region being increasingly explored. With twenty mobile observers in the low bias setting, the impact of the observer bias is negligible on the prediction performance of the uncorrected model and it outperforms the effort-corrected model (Fig \ref{fig:APV_DGM}). Conversely, in the low observer bias setting, the MSPE from the effort-corrected approach is found to increase with the number of observers. Here, increases in the squared prediction bias dominate any possible reductions in the variance of predictions. In the high bias setting, the reverse relationship is seen. Here, reductions in prediction variance offered by the increased number of observers offsets any increases in the squared prediction bias. Fig S\ref{fig:Bias_DGM} shows that in this setting, the variability in the estimates of the UD center decreases substantially as the number of observers increases. 

We demonstrate our claims made above in an additional simulation study explained in depth in the supplementary material. In it, we change two simulation settings. First, we increase the speed of the movements across each time step to reduce the autocorrelation between the encounters. This moves the simulation from the pseudo continuous-time encounter setting to a more discrete-time setting. Second, we fit an additional effort-corrected model that directly accounts for the overlap between the observers' fields-of-view. This `overlap-corrected model' is found to completely eliminate the estimation bias of the UD center (Fig S\ref{fig:Bias_overlap}). Interestingly however, no change in the MSPE is witnessed relative to the previous effort-corrected model (Fig S\ref{fig:APV_overlap}). Thus it appears that the bias reduction from the overlap-correction approach comes at a cost of an increased variance of the UD predictions. Both effort-corrected models outperform the uncorrected model with respect to MSPE across all levels of observer bias.



In summary, it appears that the benefits of effort-correction can be attained when little is known about the precise nature of the observer effort. As long as the animal's UD remains reasonably constant throughout the trips, crude attempts at effort correction appear to be better than ignoring effort in most settings. Furthermore, the path integrals of observers' fields-of-view appears control for effort reasonably well. When the degree of observer bias is expected to be high, as is expected in our case study, it appears that this form of effort-correction can lead to dramatic improvements in predictive performance, without the need to consider observer overlap or explicitly model the animal movement. 


\section{Application to empirical data}

\subsection{Special considerations required for our motivating problem}

To demonstrate the utility of our modeling framework, we apply it to the southern resident killer whale (SRKW) data. We partition the temporal domain (May - October) $\cal{T}$ into months $T_l : l \in \{\textnormal{May},...,\textnormal{October}\}$ and assume the intensities and hence the UDs of the pods are constant within each month. We denote the day as $d \in \{1,...,N_{T_l}\}$ and the year as $y \in \{2009, ..., 2016\}$. We assume that no changes to the UDs occur between 2009-2016. Our motivating dataset contains several special features that require careful consideration.

As mentioned in Section 2.2, the pod identities (J, K, or L) of the sightings can be considered known. We denote the pod identity for a sighting with the discrete mark $m$ and we consider the pods as our `individuals'. Pods are often found swimming together in `super-pods'. We break up sightings of super-pods into their individual components. For example, if a sighting of super-pod JK is made (i.e. J and K are found together), then we record this as a sighting of pod J and a sighting of pod K and ignore the potential interaction.

The data are heavily autocorrelated. Sightings are often made of the same pod in quick succession, and the locations of whale pod sightings are shared between whale-watch operators. In fact, once a pod has been sighted, it is rarely lost by the tour operators for the remainder of the day. To remove the autocorrelations, we consider only the first sightings per day of each pod, discarding all repeated sightings made within a day. Importantly, for each day and for each pod, we also discard all predicted effort that occurs after the initial sighting. Because whales move quickly relative to $|\Omega|$, an overnight window between sightings is sufficient to remove the autocorrelation between sightings. Next, we estimate the cumulative monthly observer effort from all observers. The effort is summed across the 8 years. 
 
\subsection{Incorporating the observer effort from the DFO data}

The daily GPS tracklines of the DFO vessel prior to each initial SRKW sighting are used to approximate the DFO's observer effort. The GPS data is irregular, with a typical resolution of around 15 seconds. We predict the locations at regular 30 second intervals using a continuous-time correlated random walk model fit to each trip using the crawl package \citep{crawl1, crawl2}. We denote the approximate locations and effort as $\xi_{DFO}(t, y, d)$ and $E^{obs}_{DFO}(\textbf{s}, y, T_l, m)$ respectively. Next, we count up the number of predicted points that fall into a set of polygonal regions $A_i$ used to approximate (1). Thus, we assume that at each 30 second interval, the observer's field-of-view $\phi_{DFO}(t, y, d)$ is uniform throughout the $A_i$ that contains the vessel. Thus, $\int_{A_i} E^{obs}_{DFO}(\textbf{s}, y, T_l, m) d\textbf{s} \approx \sum_d \int_{T_{l}} \mathbb{I}\{\xi_{DFO}(t, y, d) \in A_i \} dt$. The $A_i$ are approximately circular with radius 2.6km (see Fig \ref{fig:mesh}). Note that we only have the location of the vessel during encounters. Results from the simulation study suggest that these steps are unlikely to significantly impact the analysis, given the large number of boat tracks available, the large $\Omega$, and given that our assumed maximum detection range of 2.6km for $\phi_{DFO}(t)$ is likely not orders of magnitude form the truth. Effort is scaled into units of hours.



\subsection{Estimating the whale-watch observer effort}

To incorporate the observer effort from the whale-watch vessels, we build a stochastic emulator of the cumulative `boat-hours' spent in each of the integration points $A_i$ by the whale-watch companies for each day, month, and year under study. We refer to the cumulative pod-specific monthly whale-watch observer effort intensity as $E^{obs}_{WW}(\textbf{s}, y, T_l, m)$. Because the whale-watch sightings are not linked to a specific vessel, we assume throughout that the observer efficiencies across the whale-watch vessels are constant. We do not adjust for overlap between the fields-of-view of the vessels. The density of boats within the study region is expected to be far smaller than it was in the simulation study with twenty vessels and the degree of observer bias is very high, suggesting that the results should be accurate.  Note that the assumptions made on $\phi_{WW}(t, y, d)$ match those of $\phi_{DFO}(t, y, d)$. 

For each day and for each pod, we first record the number of hours into the operational day at which the initial discoveries were made. We denote this $\tau$. We assume that the daily operational period for the whale-watch companies is 9am - 6pm \citep{seely2017soundwatch}, thus $\tau \in [0, 9]$. As an example, suppose that on a given day, pods J and K were both sighted at 12pm and pod L was never sighted. Then $\tau$ would be recorded as 3 hours for pods J and K and 9 hours for pod L. To account for the changing effort throughout the day, we use the numbers of vessels with whales by hour of day reported by Soundwatch to estimate a cumulative distribution function $F_{E}(\tau)$. For an initial sighting of a pod made $\tau$ hours into the day, $F_{E}(\tau)$ represents the fraction of total whale-watch effort spent prior to that sighting. 


Let $\left(\tau_{m,y,T_l,d}\right)_{d=1}^{N_{T_l}}$ denote the number of hours after 9am when the first sighting of pod $m$, in year $y$, in month $T_l$ and on day $d$ occurs. Under the assumption that an overnight window removes the autocorrelation between the SRKW locations, the fraction of total WW observer effort spent prior to the initial sightings of pod $m$ in a given month/year is: 
\begin{align}
    \frac{1}{N_{T_l}} \sum_{d=1}^{N_{T_l}} F_{E}(\tau_{m,y,T_l,d}).\label{fractiontotaleffort}
\end{align}


Next, we need to estimate the maximum possible number of boat hours of observer effort for each year, month, and day. We denote it $E_{WW}(y, T_l, d)$. We will then multiply the year, month sums $ E_{WW}(y, T_l) = \sum_{d=1}^{N_{T_l}} E_{WW}(y, T_l, d)$ by the fraction (\ref{fractiontotaleffort}). The result will be an estimate of the observer effort associated with the initial sightings. This requires some strong assumptions that are detailed in the supplementary material, including that the average spatial distribution of the whale-watch boat observer effort is constant throughout the day. 

Soundwatch reports on: the number of active whale-watch ports per year, the maximum number of trips departing each day from each port, the changing number of daily trips across the months, and the duration (in hours) of the trips from each port. We also download wind-speed data and ask various operators for their operational guidelines on cancellations due to poor weather/sea state. We then remove days considered `dangerous'. Given the large sources of uncertainties associated with estimating the above quantities, we formulate probability distributions to appropriately express the uncertainties with each of our estimates. These probability distributions form the backbone of our stochastic emulator of $E_{WW}(y,T_l,d)$.  

To estimate the spatial distribution of the observer effort, we estimate how many boat hours could fall in each of the integration points $A_i$ per month and year. Estimates of maximum travel ranges from each port are obtained, considering land as a barrier. Typical vessel routes from the whale-watching companies are established through: private communications with the operators, Soundwatch reports, the operators' flyers and websites. Combining these together, we then formulate plausible effort fields from each port by hand using GIS tools. 

We denote $E_{WW}(\textbf{s}, y, T_l)$, the maximum possible observer effort intensity for year $y$, month $T_l$ and at location $\textbf{s} \in \Omega$. It is subject to the following constraint:
\begin{align}
\int_\Omega E_{WW}(\textbf{s}, y, T_l) \textrm{d}\textbf{s}  &= \textrm{Total possible WW boat hours in year } y\textrm{, month } T_l  \nonumber
 \\
&= E_{WW}(y, T_l). \nonumber
\end{align}

Our estimate of the pod-specific monthly whale-watch observer effort surface associated with our initial daily sightings is:

\begin{align}
E^{obs}_{WW}(\textbf{s}, T_l, m) &= \sum_{y = 2009}^{2016} E^{obs}_{WW}(\textbf{s}, y, T_l, m) \\
E^{obs}_{WW}(\textbf{s}, y, T_l, m) &= E_{WW}(\textbf{s}, y, T_l) \times \sum_{d=1}^{N_{T_l}} F_{E}(\tau_{m,y,T_l,d}) \nonumber. 
\end{align}

\subsection{Combining effort surfaces} 

Due to their spatially disjoint observer efforts (Fig \ref{fig:WWandDFOsightings}), almost no spatial overlap exists between the two sources of sightings: presence-only sightings (reported by the whale-watch operators) and the presence-absence data (recorded from the DFO boat survey).  Consequently, under our LGCP framework, any intercept term added to $\textbf{w}_1(\textbf{s}, T_l, m)$ for capturing the relative observer efficiencies between the two observer types will not be estimable due to confounding with the spatial field $Z(\textbf{s}, T_l, m)$. 

Since both observer types involve similarly-sized vessels, we make the assumption that the efficiencies across the two observer types are identical. Thus, we simply sum the two observer effort layers to get the total observer effort:  

\begin{align}
    E^{obs}_{Total}(\textbf{s}, T_l, m) = E^{obs}_{WW}(\textbf{s}, T_l, m) + E^{obs}_{DFO}(\textbf{s}, T_l, m)
\end{align}


Large uncertainties surround our estimates of the whale-watch observer effort, and failing to account for these uncertainties could lead to over-confident inference. We produce $G$ Monte Carlo samples of the effort field $E^{obs}_{WW, g}(\textbf{s}, T_l, m) : g \in \{1,...,G\}$. For each sampled observer effort field, we then fit the LGCP model and sample once from the posterior distributions of all the parameters and random effects. These new posterior distributions will help account for the uncertainty in observer effort, so long as $G$ is chosen sufficiently large to reduce the Monte Carlo error. We choose $G=1000$.



\subsection{Model selection}

We propose and fit several candidate models of increasing complexity for analysis. We fit the models using the R-INLA package with the SPDE approach \citep{rue2009approximate, lindgren2011explicit, lindgren2015bayesian, R}. All models use the estimated observer effort field $E^{obs}_{Total}$, with no detectability or observer effort covariates used (i.e. $p_{det} \equiv 1$ and $\lambda_{eff} \equiv E^{obs}_{Total}$). Model candidates start from the simplest complete spatial randomness model. This assumes that conditioned on observer effort, encounter locations for each pod and month arise from a homogeneous Poisson process. They finish with models for $\lambda_{true}(\textbf{s}, T_l, m)$ which include: covariates, temporal splines, and Gaussian (Markov) random fields with separable spatio-temporal covariance structures. To avoid excessive computation time, we perform model selection on a single realisation of our observer effort field. Then, for our `best' model, we propagate the uncertainties with observer effort through to the results via the Monte Carlo approach.  

\begin{center}
    \begin{figure}
        \centering
        \includegraphics[scale=0.4]{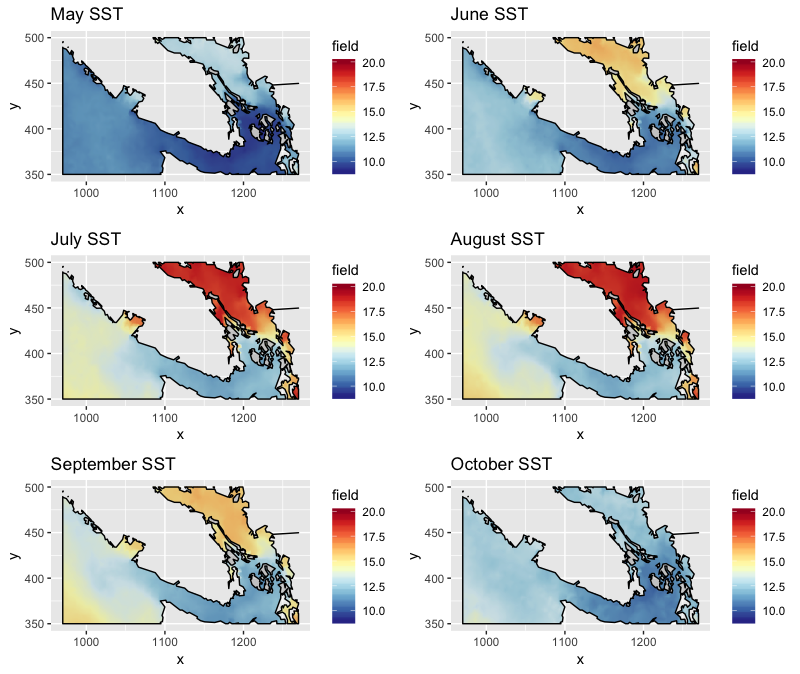}
        \includegraphics[scale=0.4]{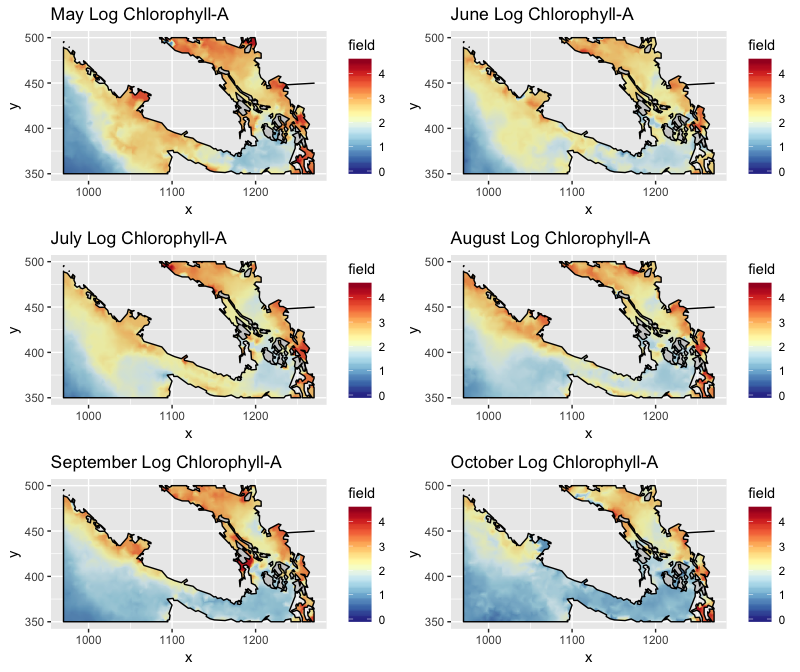}
        \caption{Plots showing the average monthly sea-surface temperatures (top 6) and the natural logarithm of chlorophyll-A concentrations (bottom 6). The averages have been taken over the years 2009-2016. }
        \label{fig:covariates}
    \end{figure}
\end{center}

We explore two space-time covariates and one spatial covariate: sea-surface temperature (SST), chlorophyll-A (chl-A), and depth. Covariates were downloaded from the ERDDAP database \citep{ERDDAP_data}, with the monthly composite SST and chl-A rasters (Fig \ref{fig:covariates}) extracted from  satellite level 3 images from the Moderate Resolution Imaging Spectroradiometer (MODIS) sensor  onboard  the  Aqua  satellite (Data  set  ID's:  erdMH1sstdmday and erdMH1chlamday respectively). We compare the two types of hierachical space-time centering seen in \citet{yuan2017point}. 

We also explore random fields and splines. Including these adds a substantial amount of complexity to the model. To avoid over-fitting the data, we start with the simplest models without random effects, and iteratively increase the complexity of the model in a stepwise manner. To choose the `best' model, we use the Deviance Information Criterion (DIC). This trades-off the goodness-of-fit of the model with a penalty for the model's complexity \citep[see][]{spiegelhalter2002bayesian}. The candidate models that do not contain random fields or splines are equivalent to MAXENT models, a commonly used SDM method \citep{renner2013equivalence}. Thus the DIC values of the models allow for comparisons to be made between the commonly used MAXENT models and our proposed LGCP model. 

We also conduct posterior predictive checks on the candidate models \citep{gelman1996posterior}. In particular, we assess the ability of the models to accurately estimate the total number of first sightings of each pod, per month. We also assess the models' abilities to suitably capture the spatial trend by comparing the observed number of sightings falling within each region $A_i$ with their model-estimated credible intervals.

\subsection{The final selected model}

The final `best' model, as judged by DIC and posterior predictive check assessments, includes a spatial random field shared across the three pods, a spatial field unique to pod L, pod-specific temporal effects (as captured by second-order random walk processes), SST, and chl-A. Both covariates were space-time centered. Depth was omitted as its inclusion led to numerical instabilities due to high multicollinearity. Details of all the models are in the supplementary material (see Table S\ref{Table:DIC}).   

\begin{figure}[ht!]
        \centering
        \includegraphics[scale=0.38]{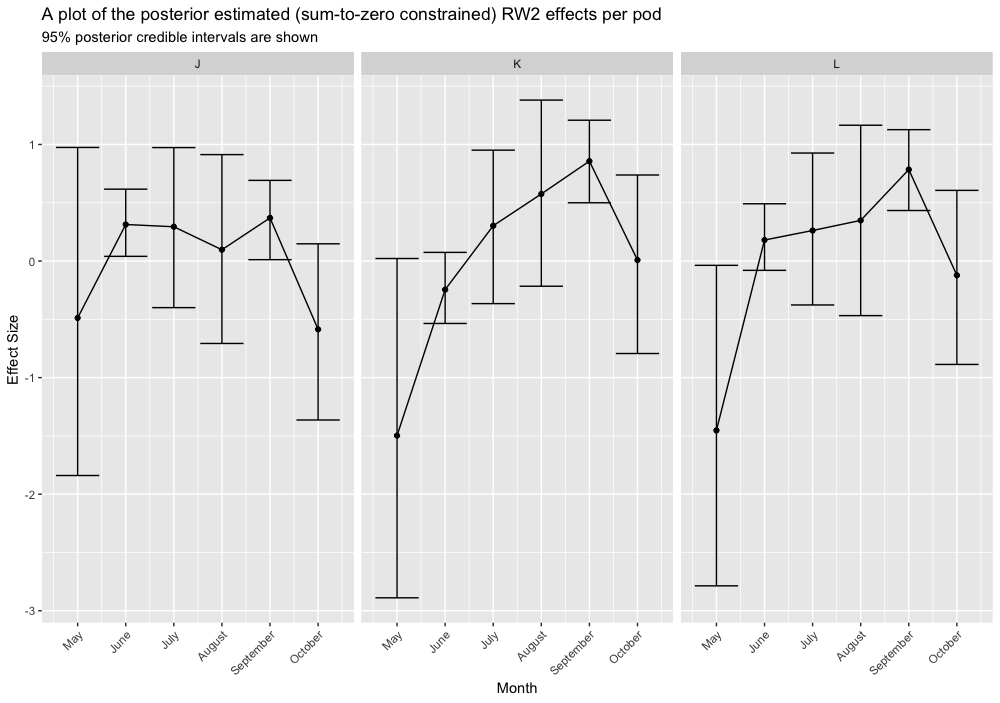}
        \caption{A plot showing the posterior mean and posterior 95\% credible intervals of the pod-specific (sum-to-zero constrained) random walk monthly effect from the `best' model with Monte Carlo observer effort error included.}
        \label{fig:RW2}
    \end{figure}

The importance of including a random field unique to pod L implies that pod L exhibits different space use compared with J and K. This result is in agreement with \citep{hauser2007summer}. No unique spatial field for pod J or K was found to significantly improve the model. The pod-specific random-walks reflect the different times the pods arrive and leave the area of interest. For example pod J is found to remain in the area of interest across the months, whereas pods K and L are found to have lower intensity in May relative to September (Fig S\ref{fig:RW2}). This is in agreement with \citet{ford1996killer} 104 pp. 

Finally, we use the the causal DAG shown in Fig \ref{fig:causal_dag} to display our assumptions about the `best' model. The first assumption is that we can accurately emulate observer effort $\lambda_{eff}(\textbf{s}, T_l, m)$ with $E^{obs}_{Total}(\textbf{s}, T_l, m)$ and that no unmeasured strong predictors of effort $\textbf{w}^\star(\textbf{s}, T_l, m)$ exist. Large residual spatio-temporal correlations caused by $\textbf{w}^\star(\textbf{s}, T_l, m)$ would be erroneously captured in the spatial fields for $\lambda_{true}(\textbf{s}, T_l, m)$, leading to estimates of pod intensity to remain confounded by effort. The second assumption is that no path in the DAG exists between the environmental covariates, measured or unmeasured (i.e. $\textbf{x}(\textbf{s}, T_l, m)$ nor $\textbf{x}^\star(\textbf{s}, T_l, m)$), and the effort $\lambda_{eff}(\textbf{s}, T_l, m)$. This would also lead to pod intensity estimates to remain confounded by effort. For the estimates of the species-environment effects $\boldsymbol{\beta}$ to be accurate, we need to assume that no unmeasured environmental covariates $\textbf{x}^\star(\textbf{s}, T_l, m)$ exist. This assumption is unlikely to hold. The choices driving the movements of the SRKW are likely far more complex than explained by the two covariates alone and unmeasured environmental factors $\textbf{x}^\star(\textbf{s}, T_l, m)$ are likely to interact with both $\textbf{x}(\textbf{s}, T_l, m)$ and $\lambda_{true}(\textbf{s}, T_l, m)$ causing confounding. However, the presence of $\textbf{x}^\star(\textbf{s}, T_l, m)$ should not impact our ability to predict $\lambda_{true}(\textbf{s}, T_l, m)$, since any strong residual autocorrelations due to $\textbf{x}^\star(\textbf{s}, T_l, m)$ should be captured by $Z(\textbf{s}, T_l, m)$.

\subsection{Displaying the results}

Large uncertainties surround our estimates of the SRKW intensity (i.e.\ encounter rate). Side-by-side maps of posterior mean and posterior standard deviation can prove challenging to interpret, making it difficult to determine regions of `high' intensity. Instead, by using a large number of posterior samples ($G$) from our model, we are able to compute exceedance probabilities and then clearly display both point estimates with uncertainty in a single map, called an exceedance map.
 
Exceedance maps display the posterior pointwise probabilities that the value of a random surface evaluated across a regular lattice grid of points exceeds a chosen threshold. For our application, we are interested in identifying regions of high whale intensity. As such, our maps will display the posterior pointwise probabilities for month $t$ that the pod-specific intensity $\lambda_{true}(\textbf{s}, T_l, m)$, at location $\textbf{s}$, lies above a chosen intensity threshold value. Here, we choose the 70th percentile of that pod's intensity. Hotspots are then identified by displaying only the points that have a posterior pointwise probability above a probability threshold. We choose a probability threshold of 0.95, which represents areas where the model predicts with at least a probability of 0.95 that the posterior intensity is in the top 30\% of values for that pod. Such maps simultaneously present our point (i.e. `best') estimates whilst also reflecting the uncertainties surrounding these estimates. For example, regions predicted to have a high encounter rate, but also a large uncertainty (e.g. regions rarely visited, but where a few encounters were made), will no longer appear in these exceedance plots.

\begin{figure}[ht!]
        \centering
        \includegraphics[scale = 0.119]{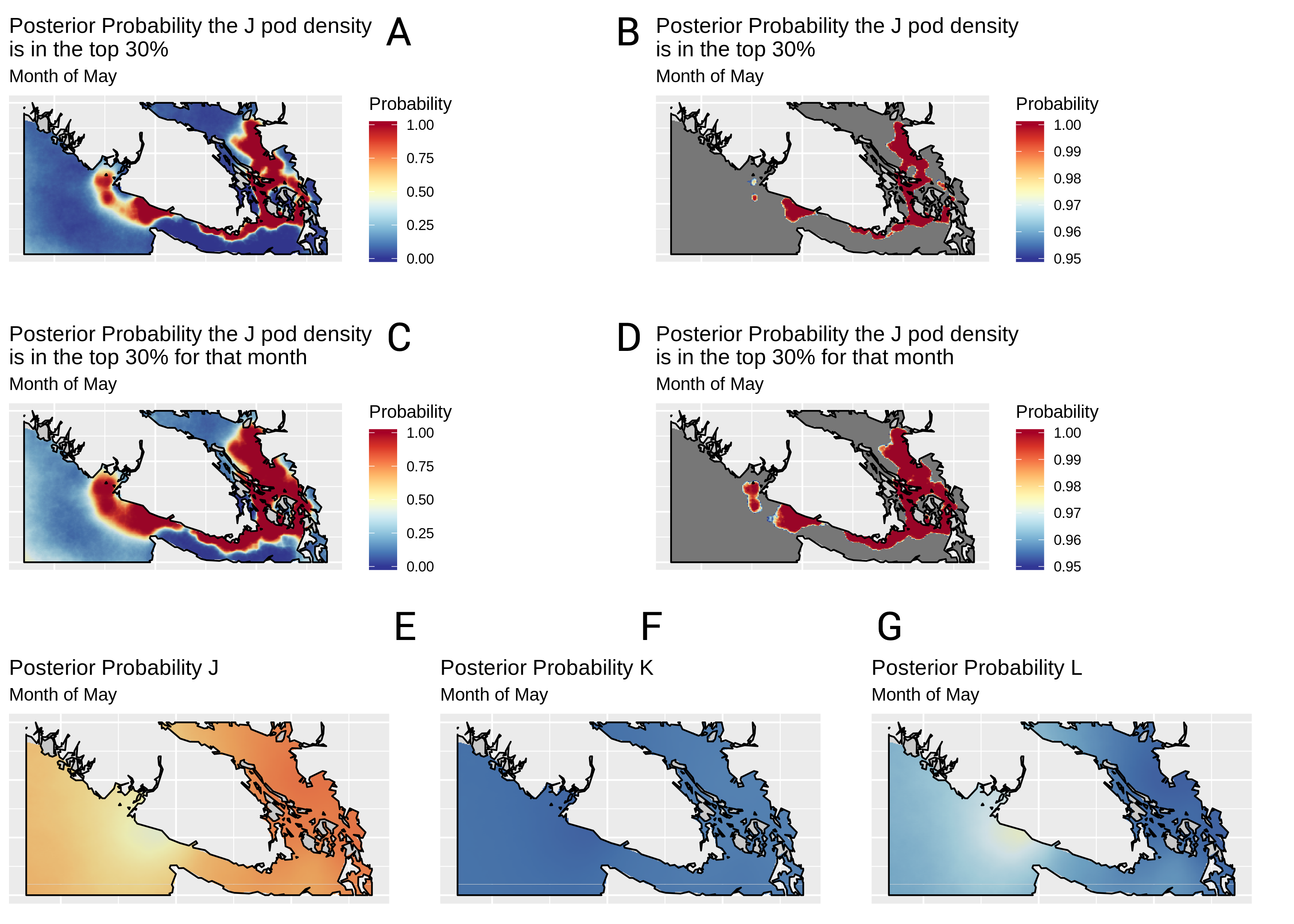}
        \caption{ A series of plots demonstrating the different types of plots possible under our modeling framework. Panels A and B show the posterior probability that J pod's intensity across the region takes value in the upper 30\% in the month of May. Panel A shows the raw probabilities, while the Panel B has a minimum probability threshold of 0.95. Panels C and D are same, however the upper 30\% exceedance value is defined separately for each month. Panels E, F, and G show the posterior probabilities that a sighting made at a given location in May contains pods J, K and L respectively. All results are shown for the `best' model with Monte Carlo observer effort error.}
        \label{fig:Combinedplots}
    \end{figure}

For demonstration, we explore regions that our model confidently predicts to have a high J-pod intensity $\lambda_{true}(\textbf{s}, T_l, m)$ during May. These correspond to hotspots of their UD. Panel A in Fig \ref{fig:Combinedplots} shows the posterior probability that the J-pod intensity in May lies in the top 30\% of values. The plots show clear hotspots in J-pod's May intensity in the West of the region and in inshore waters. We repeat the plot, but now colour all pixels grey for which a posterior probability of exceeding the 70th percentile value is below 0.95. This helps to differentiate the regions of interest that we are most confident about (Fig \ref{fig:Combinedplots} B). If we change the upper exceedance value to be the 70th percentile value for the month of May only, rather than across all months, the regions of interest are larger (Fig \ref{fig:Combinedplots} C-D).


Pod-probability maps can identify the core areas within $\Omega$ associated with each pod and month. For a chosen month $T_l$ and pod $m$, we define its `core area' to be a region $D_{T_l,m} \subset \Omega$ such that if an encounter is made within $D_{T_l,m}$ during $T_l$, there is a `high' probability that it is of pod $m$. Under our multi-type LGCP framework, because we can fix observer effort, we are able to compute the posterior probabilities that an encounter made at a given location and month contains a specific pod (see equation (\ref{eq:probobskequation})). For May, we display the posterior probabilities that an encounter made at location $\textbf{s} \in \Omega$ contain pod J, K and L respectively in panels E, F and G in Fig \ref{fig:Combinedplots}. It is apparent that in May, pod J is most likely to be encountered, in agreement with \citep{ford1996killer}.



\begin{figure}[ht!]
        \centering
        \includegraphics[scale = 0.8]{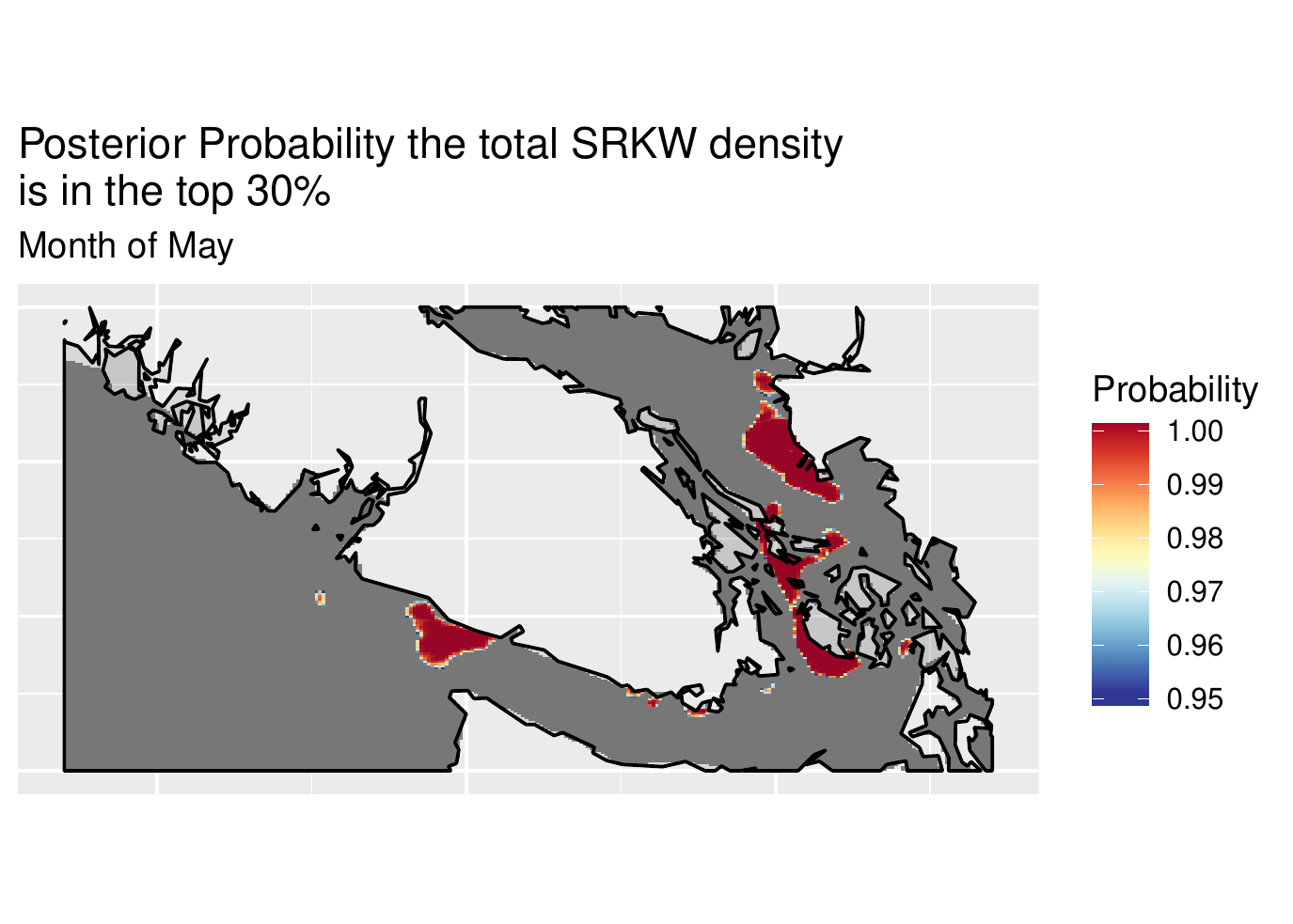}
        \caption{A plot showing the posterior probability that the sum of the three pod's intensities across the region takes value in the upper 30\% for the month of May. The 30\% exceedance value is computed across all the months. Shown are the probabilities of exceedance, with only the probabilities greater than 0.95 displayed. Results shown are for the `best' model, adjusted for Monte Carlo observer effort error.}
        \label{fig:Posterior70exceedance90_ALL}
    \end{figure}

\begin{center}
    \begin{figure}[ht!]
        \centering
        \includegraphics[scale = 0.5]{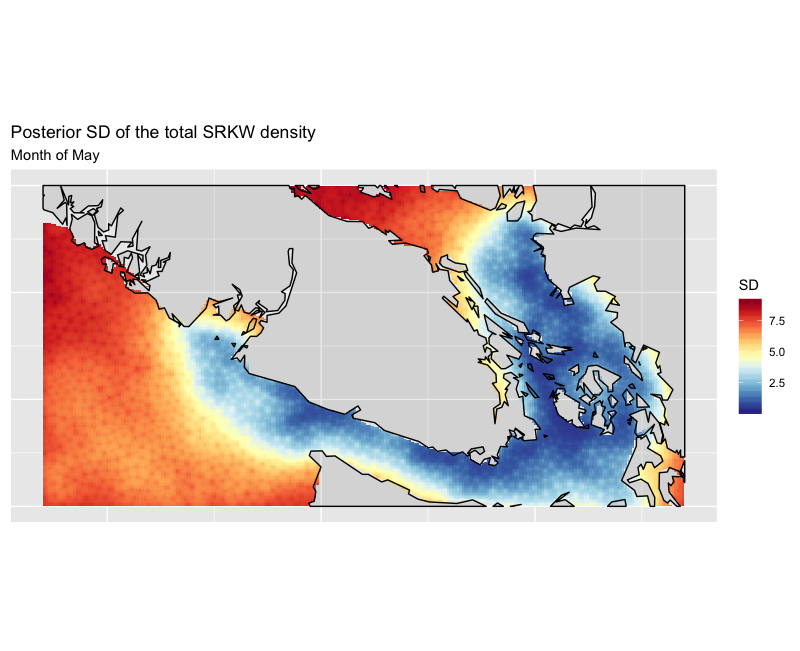}
        \caption{A plot showing the posterior standard deviation of the sum of the SRKW intensities for the three pods, for the month of May. The qualitative behaviour is almost identical across all pods and across all months, so we omit them. Results shown are for Model 8 with MC observer effort error. Note that the computational mesh is visible in the plot as we linearly interpolated the standard deviations from the computational mesh vertices to the pixel locations, instead of approximating the full posterior distributions at each pixel location. This was done to reduce computation time.}
        \label{fig:PosteriorSD}
    \end{figure}
\end{center}

When the sightings of every individual from the target population are available, one can sum the individuals' intensities and then normalise to create estimates of the population-level distribution. Maps of the population's distribution may be especially useful for conservation purposes. See for example Fig S\ref{fig:Posterior70exceedance90_ALL}, where we fix the upper value to exceed as the 70th percentile value of the sum of the three pod's intensities across all months. We assume that individuals strictly swim in their pods, and that each pod is a single unit of identical size. We do not scale the pod-specific intensities by their group sizes. Thus, the intensity represents the expected number of encounters of any pod per boat hour of effort. The effort $E^{obs}_{Total}(\textbf{s}, T_l, \textbf{m})$ is estimated to be nonzero throughout most of the region $\Omega$, with two exceptions. The first is the region in the very top of $\Omega$, to the West of Vancouver. The second is in the Northwestern corner of $\Omega$. These regions were never visited and so little can be said about the true SRKW intensity in these regions. This is reflected in the very large posterior standard deviations shown there in Fig S\ref{fig:PosteriorSD}.

\section{Discussion}

We have built upon the recent developments made in the species distribution modeling literature and presented a general framework for estimating an individual's utilization distribution (UD). In addition, we have shown that these estimates can be combined to form the spatio-temporal distribution of a species or group. We demonstrated its use by identifying areas frequently used by an endangered ecotype of killer whale. Using the methodology, data from multiple observers, and data of varying quality and type, may all be combined to jointly estimate the spatio-temporal distribution. Crucially, high-quality survey data can be combined with low quality opportunistic data, including presence-only data. Data types compatible for modeling with this framework extend beyond those seen in this motivating example. Log-Gaussian Cox processes (LGCPs) have the unifying feature of providing a base model for deriving the likelihoods of many of the commonly found data, including presence-only, presence-absence, site occupancy, and site count data \citep{miller2019recent}. Such data fusion can improve the spatial resolution and statistical precision of estimates of the spatio-temporal distribution of species \citep{fithian2015bias,koshkina2017integrated}.

However, including presence-only data requires knowledge about the observer effort, either directly (e.g. GPS records) or through a set of strong predictors (e.g. distance from the nearest road). In either case, we show that approximating the observer effort by either computing or modeling the path integrals of the observers' fields-of-view can be a relatively straightforward and successful approach. Furthermore, results from our simulation study suggest that only crude estimates of the observers' fields-of-view are required, and that substantial improvements in the accuracy of UD predictions can be attained when the degree of observer bias is high. Furthermore, these improvements are still seen in settings where substantial overlap exists between the observers' fields-of-view and where the size of the study region is small. A fundamental assumption of our work was that the utilization distributions of the individuals were stationary throughout known time intervals. This greatly simplified the task of estimating the observer effort. If this stationarity assumption is unsuitable and the UDs evolve continuously through time, then observer effort needs to be known or estimated on a continuous time scale too. Estimating unknown effort in continuous time from a set of covariates will likely prove to be a challenge.

While the mathematical theory underpinning the LGCP may appear challenging to many researchers, the application of these models is widely applicable. Recent developments in spatial point process R packages \citep{R}, such as spatstat \citep{baddeley2014package} and inlabru \citep{inlabru} facilitate their computation. Inlabru requires only basic knowledge of R packages such as sp \citep{sp1, sp2}, rgeos \citep{bivand2013rgeos}, and rgdal \citep{bivand2015rgdalpackage}. Pseudo-code is supplied in the supplementary material to show how a dataset with a combination of distance sampling survey data and opportunistic presence-only data could be analysed using this modeling framework. Joint models are fit and sampled from using only 7 function calls, emphasising the applicability of the framework across a wide range of disciplines.

A biology-focused companion paper is currently underway, using the final model outputs to explore SRKW habitat use and how it varies in this region across pods and summer months. Importantly, it will compare and contrast habitat use based on traditional opportunistic sightings data analyses and for the first time present relative SRKW habitat use across the entire extent of SRKW critical habitat in Canadian Pacific waters together with estimates of confidence. Thus the models developed in this paper will play an important role in planning future SRKW conservation efforts and highlighting regions of ecological significance.

\section{Acknowledgements}
 We would like to thank the DFO, BCCSN, The Whale Museum and NOAA for access to sightings databases. We thank Jason Wood (SMRU), Jennifer Olson (The Whale Museum) and Taylor Shedd (Soundwatch) for their detailed insight into the operations of the whale-watch industry. Additionally we would like to thank Eagle Wing Whale \& Wildlife Tours for their substantial help with developing the effort layer for Victoria. This message of thanks extends to various other whale-watch companies who also assisted with the process of estimating the observer effort. Finally, we would like to thank Jim Zidek for his consistent support and lively discussions throughout. MAM thanks the Canadian Research Chairs program and the Natural Sciences and Engineering Research Council. 

\bibliographystyle{imsart-nameyear}
\bibliography{SRKW_bibliography.bib}
\newpage

\section*{Supplementary Material}

\subsection{Additional theory on marked point processes}
Start with a Poisson process $Y$ on $\Omega$ with intensity $\lambda(\textbf{s})$. Next, take a probability distribution $p(\textbf{s}, .)$ on $M$ depending on $\textbf{s} \in \Omega$ such that, for $B \subset M$, $p(., B)$ is a measurable function on $\Omega$. A marking of $Y$ is a random subset of $\Omega \times M$ such that the projection onto $\Omega$ is $Y$ and such that the conditional distribution of $Y^\star$, given $Y$ makes the marks $m_y : \textbf{y} \in Y$ independent with respective distributions $p(\textbf{y}, .)$. We now have the following theorems (adapted from \citet{kingmanc}):

[Marking Theorem]
The random subset $C \subset Y^\star$ is a Poisson process on $\Omega \times M$ with mean measure $\Lambda^\star$ defined:
\begin{align}
    \Lambda^\star(C) = {\int \int}_{(\textbf{s},m) \in C} \lambda(\textbf{s}) p(\textbf{s}, \textrm{d}m) d\textbf{s}
\end{align}

[Mapping Theorem]
If the points $(Y, m_Y)$ form a Poisson process on $\Omega \times M$, then the marks form a Poisson process on $M$ and the mean measure is obtained by setting $C = \Omega \times B$ in (8):
\begin{align}
    \mu_m(B) = \int_\Omega \int_B \lambda(\textbf{s}) p(\textbf{s}, \textrm{d}m) d\textbf{s}
\end{align}

if the marks take on only $K$ different values, then the theorem specializes for the $i^{th}$ mark to:

\begin{align}
    \Lambda_i(A) = \int_A \lambda(\textbf{s}) p(\textbf{s}, \{m_i\}) d\textbf{s} \hspace{1cm} A \subset \Omega
\end{align}


\subsection{Extra results of the main simulation study}

\begin{figure}
    \centering
    \includegraphics[scale=0.7]{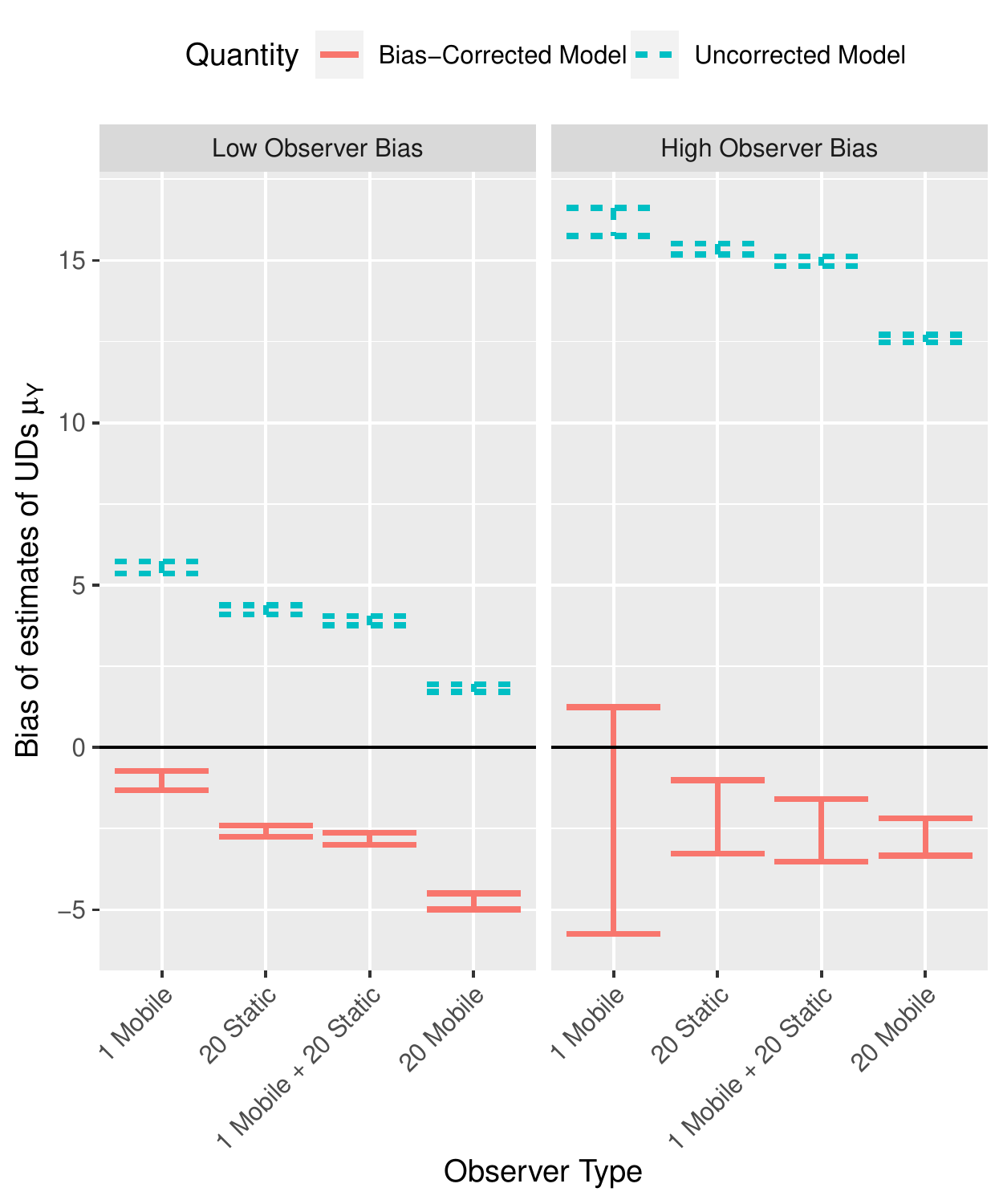}
    \caption{A plot showing the bias of the estimated y-coordinate of the animal's UD center $\mu_y$ under the bias-corrected and bias-uncorrected models vs the types of observers. From left to right are the results from one mobile observer, twenty static observers, twenty static with one mobile observers, and twenty mobile observers. The degree of observer bias is changed from low to high in the columns. The red solid lines and the blue dashed lines show the median bias along with robust intervals computed as $\pm 2c \textnormal{MAD}$ from the Bias-corrected and uncorrected models across the 100 simulation replicates respectively. The MAD has been scaled by $c = 1.48$. This ensures that the intervals are asymptotically equivalent to the 95\% confidence intervals that would be computed if the biases were normally distributed. Note that here all the analyst's assumptions correctly match the true data-generating mechanism, albeit with any overlap in the observers' efforts ignored.}
    \label{fig:Bias_DGM}
\end{figure}

\FloatBarrier
\newpage

\subsection{Details of the additional simulation study}

In the previous simulation study, we argued that two major sources of prediction bias were the autocorrelations between the encounter/non-encounter events, and the overlap between the observers' fields-of-view. We demonstrate these claims in a second simulation study. Unlike the previous simulation study, this one is designed to ensure that the encounter/non-encounter events at each time step are approximately independent of each other. This is achieved by increasing the average distance travelled by the animal at each discrete time step. This could also be interpreted as the setting where observers attempt encounters at discrete sampling times and wait for a sufficiently long amount of time between sampling times to reduce the autocorrelation between the encounter/non-encounter events. 

We simulate the movements of observers from the same stochastic differential equation model. For the animal, we change the variance of the potential function to 1 and increase the variance of the Brownian motion terms to 400. This leads to the animal moving an average distance of 23 units, compared with 1.75 and 3.5 units for the high bias and low bias mobile observers respectively. Given that each simulation trip ends when the first encounter is made, these simulation settings ensure that the autocorrelations between the encounter/non-encounter events is greatly reduced.

\begin{figure}
    \centering
    \includegraphics[scale=0.7]{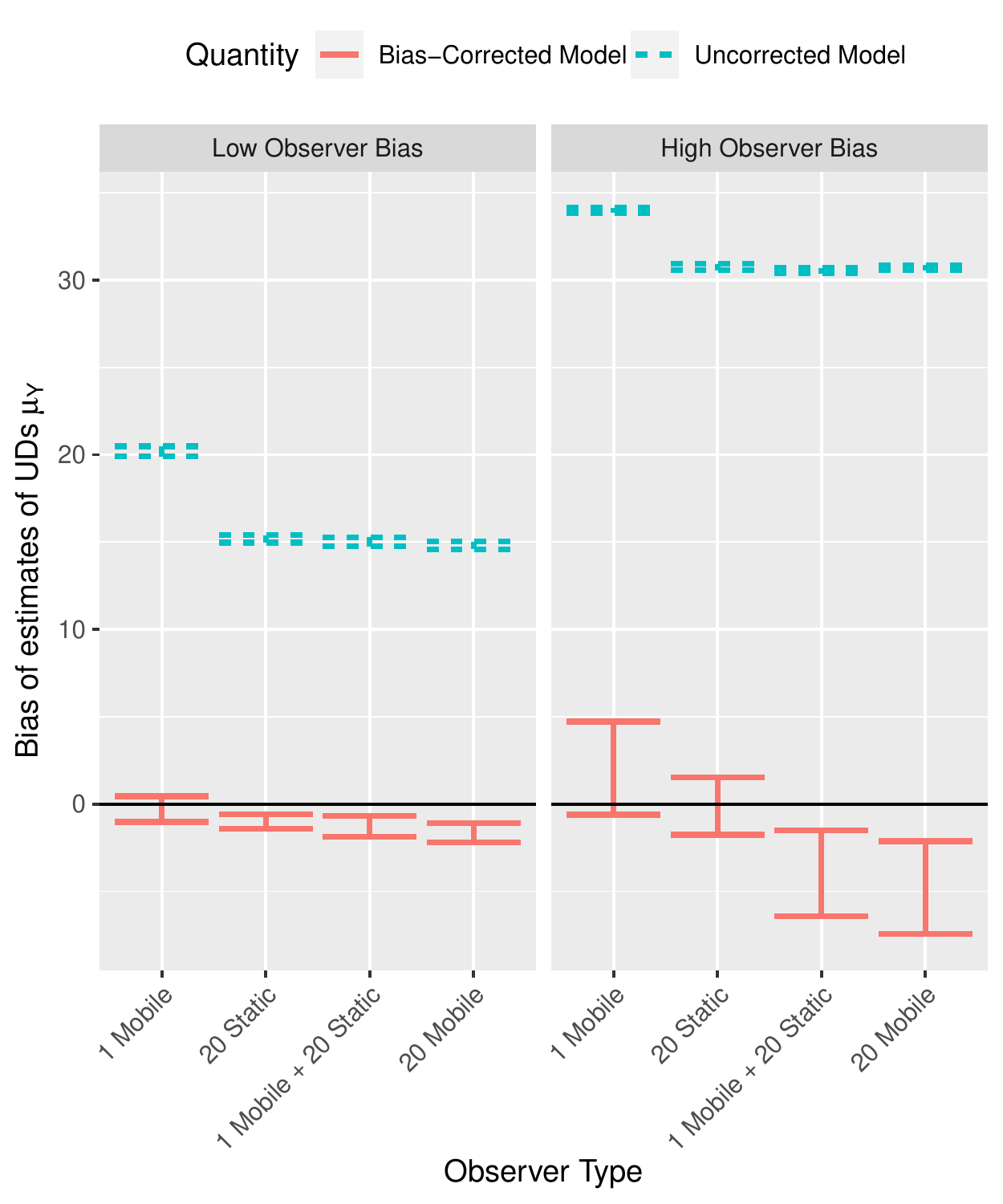}
    \caption{A plot showing the bias of the estimated y-coordinate of the animal's UD center $\mu_y$ under the bias-corrected and bias-uncorrected models vs the types of observers. The results shown here are for the second simulation study. From left to right are the results from one mobile observer, twenty static observerstwenty static with one mobile observers, and twenty mobile observers. The degree of observer bias is changed from low to high in the columns. The red solid lines and the blue dashed lines show the median bias along with robust intervals computed as $\pm 2c \textnormal{MAD}$ from the Bias-corrected and uncorrected models across the 100 simulation replicates respectively. The MAD has been scaled by $c = 1.48$. This ensures that the intervals are asymptotically equivalent to the 95\% confidence intervals that would be computed if the biases were normally distributed. Note that here all the analyst's assumptions correctly match the true data-generating mechanism, albeit with any overlap in the observers' efforts ignored.}
    \label{fig:Bias_DGM_Fastanimal}
\end{figure}

\begin{figure}
    \centering
    \includegraphics[scale=0.7]{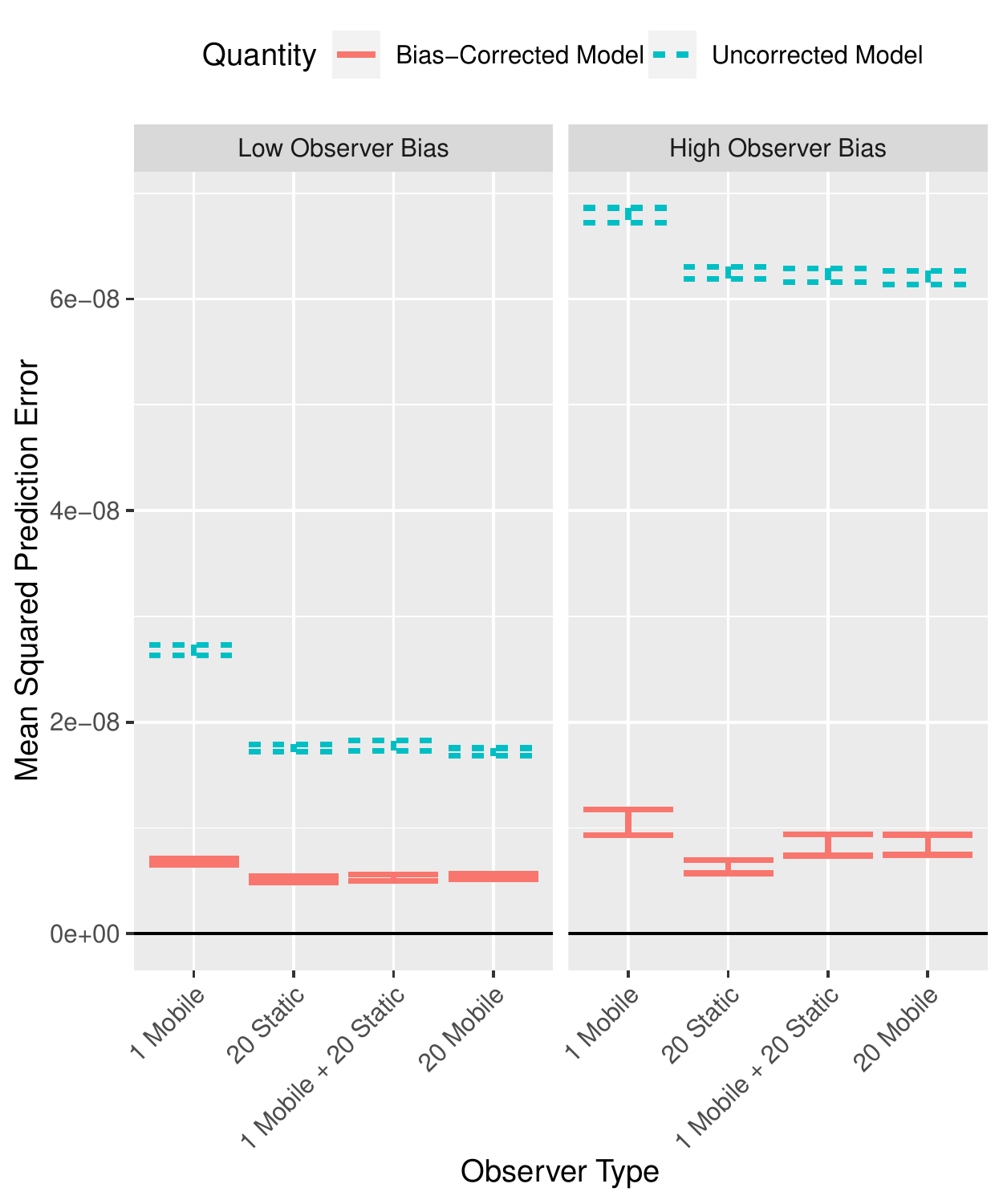}
    \caption{A plot showing the mean squared prediction error (MSPE) of the animal's UD under the bias-corrected and bias-uncorrected models vs the types of observers. The results shown here are for the \textbf{second simulation study}. From left to right are the results from one mobile observer, twenty static observers, twenty static with one mobile observers, and twenty mobile observers. The degree of observer bias is changed from low to high in the columns. The red solid lines and the blue dashed lines show the median bias along with robust intervals computed as $\pm 2c \textnormal{MAD}$ from the Bias-corrected and uncorrected models across the 100 simulation replicates respectively. The MAD has been scaled by $c = 1.48$. This ensures that the intervals are asymptotically equivalent to the 95\% confidence intervals that would be computed if the MSPE values were normally distributed. Note that here all the analyst's assumptions correctly match the true data-generating mechanism, albeit with any overlap in the observers' efforts ignored.}
    \label{fig:APV_DGM_Fastanimal}
\end{figure}

Fig S\ref{fig:Bias_DGM_Fastanimal} clearly demonstrates that a reduction in the autocorrelation between the encounter/nonencounter events leads to a reduction in the bias of the estimated UD center. Furthermore, a large increase is witnessed in the relative MSPE of the effort-corrected approach compared with the uncorrected approach. In fact, the effort-corrected approach outperforms the uncorrected approach in all settings. However, there is some remaining bias in the estimates of the UD center from the effort-corrected approach and the magnitude of this bias increases with the numbe of observers.

To demonstrate that this bias is in fact caused by overlap in the observers' fields-of-view, we implement a method for adjusting for overlap when estimating observer effort. In particular, let $p_{det}(o, \textbf{s}, t)$ denote the detection probability function for observer $o$, evaluated at space-time coordinate $(\textbf{s}, t)$. The standard bias correction approach simply estimates effort as $$E(\textbf{s}) = \sum_t \sum_{o \in O} p_{det}(o, \textbf{s}, t).$$ However, the probabilities of detection from overlapping observers do not sum. We correct for this and compute:

$$E(\textbf{s}) = \sum_t \left( 1 - \prod_{o \in O} \left(1 - p_{det}(o, \textbf{s}, t)\right) \right).$$

\begin{figure}
    \centering
    \includegraphics[scale=0.7]{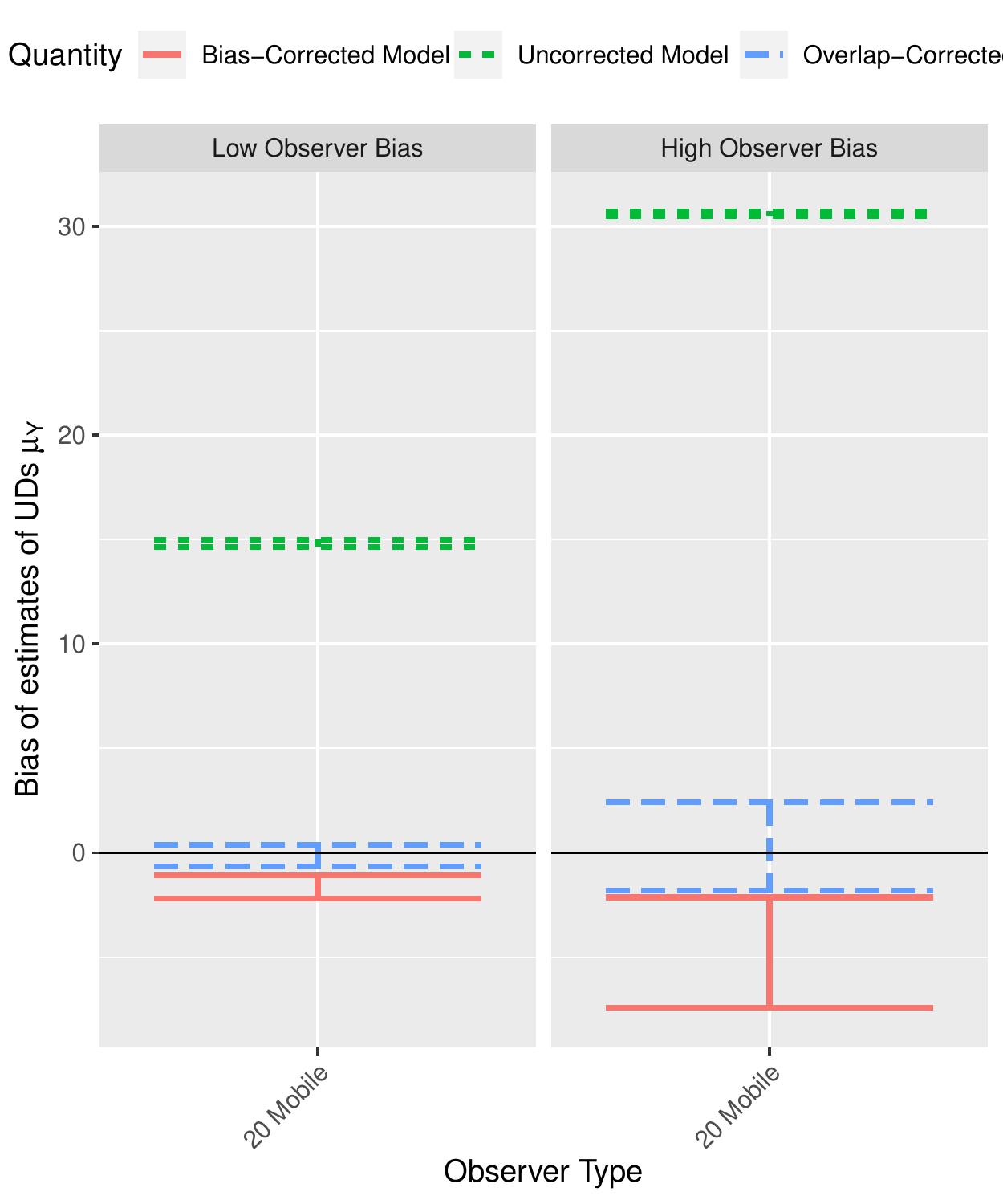}
    \caption{A plot showing the bias of the estimated animal's UD center $\mu_y$ under the bias-corrected, bias-uncorrected, and the overlap-corrected models for the twenty mobile obervers. From left to right are the results when the degree of observer bias was either low or high. The red solid lines, the blue dashed lines, and the green dotted lines show the median bias along with robust intervals computed as $\pm 2c \textnormal{MAD}$ from the Bias-corrected, uncorrected, and overlap-corrected models across the 50 simulation replicates respectively. The MAD has been scaled by $c = 1.48$. This ensures that the intervals are asymptotically equivalent to the 95\% confidence intervals that would be computed if the biases were normally distributed. Note that here the correct data-generating mechanism was assumed by the analyst.}
    \label{fig:Bias_overlap}
\end{figure}

\begin{figure}
    \centering
    \includegraphics[scale=0.7]{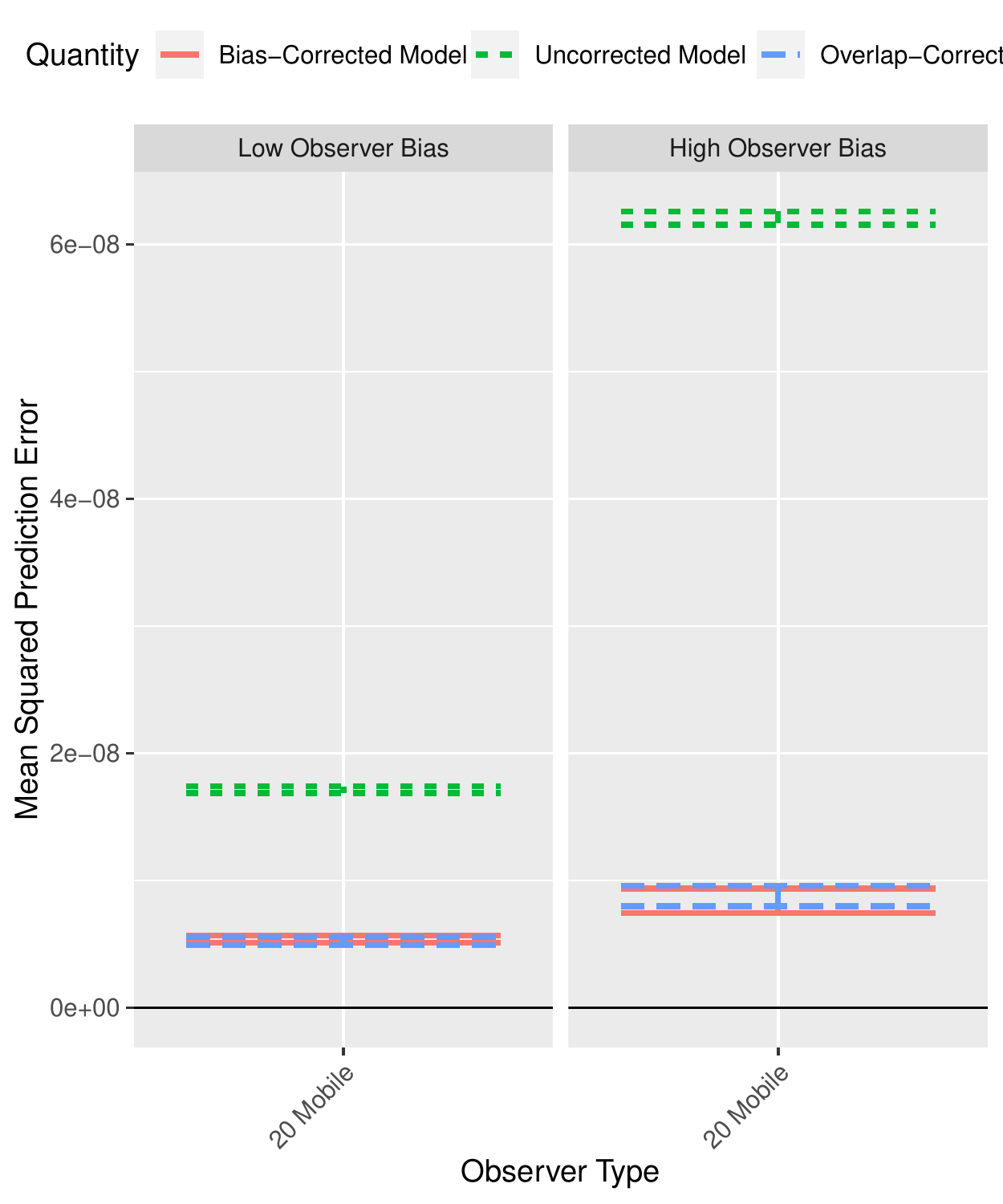}
    \caption{A plot showing the mean squared prediction error (MSPE) of the estimated animal's UD under the bias-corrected, bias-uncorrected, and the overlap-corrected models for the twenty mobile obervers. From left to right are the results when the degree of observer bias was either low or high. The red solid lines, the blue dashed lines, and the green dotted lines show the median MSPE along with robust intervals computed as $\pm 2c \textnormal{MAD}$ from the Bias-corrected, uncorrected, and overlap-corrected models across the 50 simulation replicates respectively. The MAD has been scaled by $c = 1.48$. This ensures that the intervals are asymptotically equivalent to the 95\% confidence intervals that would be computed if the MSPE values were normally distributed. Note that here the correct data-generating mechanism was assumed by the analyst.}
    \label{fig:APV_overlap}
\end{figure}

We implement this approach for 50 simulation iterations in the settings with twenty mobile observers. This setting is chosen since it suffers from the largest degree of overlap. Fig S\ref{fig:Bias_overlap} demonstrates that the overlap corrected method indeed removes the bias in estimates of the animals UD center in both the low observer bias and high observer bias settings. However, no improvement in the MSPE is seen in Fig S\ref{fig:APV_overlap}.

\FloatBarrier
\newpage

\subsection{Additional comments on the causal DAG}

\begin{figure}
    \centering
    \includegraphics[trim={0.07cm 1cm 0 0},clip,scale=0.3]{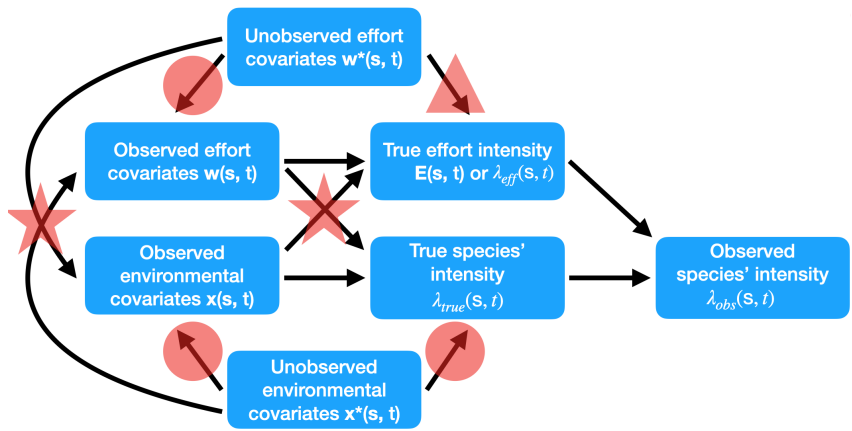}
    \caption{A plot showing the assumed causal DAG for the proposed framework with the detection probability assumed constant. An arrow between a variable set $A$ and a variable set $B$ indicates that at least one variable exists in both sets with a direct causal effect between them. The causal Markov assumption is made such that a variable is independent of its non-descendants, when conditioned on its parents \citep{hernan2010causal}. }
    \label{fig:causal_dag_Supp}
\end{figure}

Model (10) is fit to a set of observed environmental covariates $\textbf{x}(\textbf{s}, t)$ and observed effort covariates $\textbf{w}(\textbf{s}, t)$, but in general, there may exist unobserved covariates $\textbf{x}^\star(\textbf{s}, t)$ and $\textbf{w}^\star(\textbf{s}, t)$. These unobserved covariates, in conjunction with the causal paths contained in the stars, the circles, and the triangle of Fig \ref{fig:causal_dag} may cause problems. For example, the lower causal path denoted by the arrow within the red star on the left combined with the causal path within the red circle on the bottom right opens a back-door pathway between the effort intensity surface and the true species' intensity surface. This pathway passes through the unobserved environmental covariates $\textbf{x}^\star(\textbf{s}, t)$ causing estimates of $\boldsymbol{\gamma}_1^T$ to be confounded by $\textbf{x}^\star(\textbf{s}, t)$. 

A similar conclusion may be drawn by considering the upper of the two arrows within the left had star, combined with the arrow seen in the triangle. Here, estimates of $\boldsymbol{\beta}^T$ will be confounded by $\textbf{w}^\star(\textbf{s}, t)$. Further problems would occur due to the two causal paths within the right star. These would lead to $\lambda_{true}(\textbf{s}, t)$ and $E(\textbf{s}, t)$ becoming non-identifiable. The existence of a subset of covariates $\Tilde{\textbf{w}}(\textbf{s}, t)$ within $\textbf{w}(\textbf{s}, t)$ driving both $\lambda_{true}(\textbf{s}, t)$ and $E(\textbf{s}, t)$ causes neither intensity surface to be estimable. This is because only the sum of the effects of $\Tilde{\textbf{w}}(\textbf{s}, t)$ are estimable within the loglinear model (10). Thus for the true species' intensity surface to not be confounded by the effort intensity, none of the causal paths within the red stars can exist. 

Yet more problems can occur if the four causal paths within the red circles and the red triangle exist. The upper two paths lead to estimates of $\boldsymbol{\gamma}_2^T$ being confounded by unmeasured effort covariates $\textbf{w}^\star(\textbf{s}, t)$ and the bottom two paths lead to estimates of $\boldsymbol{\beta}^T$ being confounded by unmeasured environmental covariates $\textbf{x}^\star(\textbf{s}, t)$. Furthermore, the existence of the causal path in the red triangle alone may lead to estimates of $\lambda_{true}(\textbf{s}, t)$ within (10) to be confounded by $\textbf{w}^\star(\textbf{s}, t)$. This is because if a Gaussian process $Z(\textbf{s}, t)$ is included within the linear predictor for $\lambda_{true}(\textbf{s}, t)$ then any residual spatio-temporal correlations in the sightings data due to $\textbf{w}^\star(\textbf{s}, t)$ may be erroneously captured by $Z(\textbf{s}, t)$. 

Note that similar issues occur if the detection probability is not constant and a detection probability surface $p_{det}(\textbf{s}, t)$ is estimated with its own set of covariates. An extension to this causal DAG would allow for similar conclusions to be drawn. For the later case study, we assume that none of the causal paths within the stars are present.

\subsection{Deriving site occurrence and site count likelihoods}

Likelihoods for site occurrence and site count data can all be derived from the modeling framework if the true locations of the target species follow a log-Gaussian Cox process. We ignore time for notational simplicity. With $\Omega$ our study region, with a known sampled region (e.g. a transect) $A_i \subset \Omega$, and with a known or estimable observer effort captured by $\lambda_{eff}(\textbf{s}, \textbf{m})$, define the following quantity:

\begin{align*}
    \Lambda_{obs}(A_i, \textbf{m} | Z) = \int_{A_i} \lambda_{true}(\textbf{s}, \textbf{m})p_{det}(\textbf{s}, \textbf{m})\lambda_{eff}(\textbf{s}, \textbf{m}).
\end{align*}

This is referred to as the integrated observed intensity function, conditional upon knowing the Gaussian process Z(\textbf{s}). Importantly, this represents the expected number of observed sightings within $A_i$. Following \citet{hefley2016hierarchical}, we can then derive the target likelihoods. Firstly, suppose that an observer records the number of sightings made within $A_i$, denoted $N(A_i)$. Then the distribution of the number of counts, conditioned upon knowing $Z$, is: 

\begin{align*}
    [N(A_i, \textbf{m}) | Z] \backsim \textrm{Poisson}(\Lambda_{obs}(A_i, \textbf{m} | Z)).
\end{align*}

In practice, the Gaussian process is not known and thus needs to be estimated. Consequently, the above likelihood is an example of a spatial generalised linear mixed effects model (SGLMM). Multiple software packages exist to fit such models (e.g. R-INLA). Next, suppose that instead of recording the number of sightings made within $A_i$, a binary presence/absence indicator of presence (denoted $P(A_i)$) was recorded. The distribution of this indicator variable can also be derived from the conditional Poisson distribution on the counts. In particular, let $O(A_i) = I\left(N(A_i) > 0\right)$, with $I$ denoting the indicator function. Then the probability statement $P(O(A_i|Z) = 1) = P(N(A_i|Z) > 0) = 1 - \textrm{exp}[\Lambda_{obs}(A_i|Z)]$ implies the following conditional distribution on the indicator variables:

\begin{align*}
    [O(A_i, \textbf{m})|Z] \backsim \textrm{Bernoulli}(1 - \textrm{exp}\left[\Lambda_{obs}(A_i, \textbf{m}|Z)\right]).
\end{align*}

Once again, the likelihood is of the SGLMM format which can be computed using standard software packages. Note also that computing the integrated observed intensity function is critical across the likelihoods.

\subsection{Comments on preferential sampling}

In the setting of this paper, preferential sampling would be defined as a stochastic dependence between the observer effort and the underlying species intensity. An example would be a setting where observers focused their observer effort in areas with high species density, perhaps due to some prior knowledge on their likely locations. The biasing effects of preferential sampling on spatial prediction \citep{diggle2010geostatistical} and on the estimation of the mean intensity in ecological applications \citep{pennino2019accounting} have been shown. In particular, in the example above, spatial predictions in the unsampled regions would be positively biased, as would estimates of the mean intensity.    

In many situations this modeling framework will suitably adjust inference for any heterogeneous observer effort across $\Omega$, removing the biasing effects of preferential sampling. In cases where nonzero observer effort exists throughout the study region (i.e. where $E(\textbf{s}, \textbf{m}) > \hspace{0.1cm} 0 \hspace{0.1cm} \forall \hspace{0.1cm} (\textbf{s}, \textbf{m}) \in (\Omega \times M) )$, the estimation of $\lambda_{true}(\textbf{s}, \textbf{m})$ will be unaffected by preferential sampling. However, when a subregion $B \subset \Omega$ is never visited, (i.e. when $E(\textbf{s}, \textbf{m}) = 0 \hspace{0.1cm} \forall \hspace{0.1cm} (\textbf{s}, \textbf{m}) \in (B \times M) )$, the estimation of $\lambda_{true}(\textbf{s}, \textbf{m})$) within $B$ may be biased. To highlight this fact, suppose our study region $\Omega$ is split into a northern region $A$ and a southern region $B$. Suppose that the true intensity $\lambda_{true}(\textbf{s})$ takes value 2 within $A$ and value 1 within $B$. If only $A$ is visited, then without the availability of strong covariates explaining the differences across $A$ and $B$, then any model will wrongly overestimate the true intensity in $B$, namely the model will predict that $\lambda_{true}(\textbf{s}) = 1 \hspace{0.1cm} \forall \hspace{0.1cm} \textbf{s} \in B$.

To minimize the impacts of preferential sampling on any conclusions made using this modeling framework, extrapolating predictions into unsampled regions should be done with care, especially if it is believed that the intensity of observer effort may depend upon the underlying species' intensity. This is standard advice in any statistical analysis and is not a limitation unique to this framework.  


\subsection{More notes on estimating the whale-watch observer effort}

Two strong assumptions are required to allow us to multiply the total observer effort field by the fraction of total observer effort observed in a given month/year. We first assume that the expected spatial positions of the boats are constant throughout the time period of interest 9am - 6pm. We know that at the starts and ends of the days the boats will likely be closer to port. We assume however, that the whale-watch boats are travelling independently in equilibrium (represented by our estimated observer effort field $E_{WW}(\textbf{s}, T_l, y)$). 

Second, we assume that the boats are spread out throughout $\Omega$ sufficiently, such that the total observer effort from all the vessels (assumed equal) is additive. In other words, we assume the whale-watch boats are sufficiently spread out, such that their observation ranges do not overlap. In reality the whale-watch vessels often visit similar nature `hotspots' and hence traverse similar routes. As a consequence, they may travel close together at certain times. At these times, their combined observer effort may not scale linearly with the number of the boats. 

Given that we have chosen months to be our discretization of time, we must estimate the monthly observer effort across space, adding up the contributions of effort across the years of interest (2009 - 2016). 
\begin{align}
    E^{obs}_{WW}(\textbf{s}, T_l, m)) = \sum_{y = 2009}^{2016} E^{obs}_{WW}(\textbf{s}, y, T_l, m)) \nonumber
\end{align}

We define boat hours to be our unit of observer effort, kilometers to be our unit of distance and month to be our unit of time. Thus, $E^{obs}_{WW}(\textbf{s}, T_l, m)$ denotes the number of WW boat hours of observer effort, per unit area that occurred for pod $m$, at location $\textbf{s}$ and month $T_l$, summed over all the years of the study. In a similar flavour to the intensity surface, the effort surface is not really defined pointwise, but defined over regions of non-zero area as an integral. In particular, for a region $A \subset \Omega$ and month $T_l$, we define the total observer effort that occured inside $A$ in boat hours to be:

\begin{align}
    \Tilde{E}^{obs}_{WW}(A,T_l,m) = \int_A E^{obs}_{WW}(\textbf{s}, T_l, m) d\textbf{s} \label{eq:efbh}
\end{align}

For later computation of the LGCP, we approximate the stochastic integral required for the likelihood over a finite set of integration points. Thus, we are required to compute the integrals of the effort field over the integration points.

\subsection{Computational steps for approximating the likelihood}


The LGCP likelihood above (1) is analytically intractable, as it requires the integral of the intensity surface, which typically cannot be calculated explicitly. However, various methods exist for approximating this integral. We consider the approximation method from \citet{simpson2016going}. We present the spatial-only setting (i.e. $L=1$) and ignore marks for notational convenience. The results generalise easily to the spatio-temporal case with marks. First, $p$ suitable integration points are chosen in $\Omega$ with known corresponding areas $\{\Tilde{\alpha}_j\}_{j=1}^p$. Then, the first $p$ indices are defined to be the chosen integration points with the last $n$ indices chosen as the observed locations of the sightings $\textbf{s}_i \in \Omega$. Then, define $\boldsymbol{\alpha} = (\Tilde{\boldsymbol{\alpha}}^T_{p \times 1}, \textbf{0}_{n \times 1}^T)^T$ and  $\textbf{y} = (\textbf{0}_{p \times 1}^T, \textbf{1}_{n \times 1}^T)^T$. We define $\textrm{log}(\eta_i) = \textrm{log}(\lambda_{true}(\textbf{s}_i)
    p_{det}(\textbf{s}_i)
    \lambda_{eff}(\textbf{s}_i))$. We obtain:

\begin{align}
    \pi(\textbf{y} | \textbf{z}) \approx K \prod_{i = 1}^{n + p} \eta_{i}^{y_i} \rm{exp}(-\alpha_i \eta_i).\label{eq:approxlikelihood}
\end{align}  

We can see that the stochastic integral is only approximated across the first $p$ integration points, hence the name. The expected count around an integration point scales linearly with the area $\{\Tilde{\alpha}_j\}$ associated with it. This is under the assumption that for fixed intensity, doubling the area of a region, doubles the expected number of encounters occurring within the region. The problem of evaluating (\ref{eq:LGCPlikelihood}) is reduced to a problem similar to evaluating $n + p$ independent Poisson random variables, conditional on $\textbf{Z} = \textbf{z}$, with means $\alpha_i \eta_i$ and `observed' values $y_i$. This is a Riemann sum approximation to the integral. In standard software, the natural logarithm of the weights $\alpha_i$ is added as an offset in the model and equation (\ref{eq:approxlikelihood}) can be fit if one defines the minor modification that $\textrm{log} (\alpha_i)$ is defined to be zero if $\alpha_i = 0$. This is implemented as standard in the R-INLA package \citep{rue2009approximate,lindgren2011explicit, lindgren2015bayesian}.

\begin{figure}[ht!]
        \centering
        \includegraphics[width = 14cm, height=5cm]{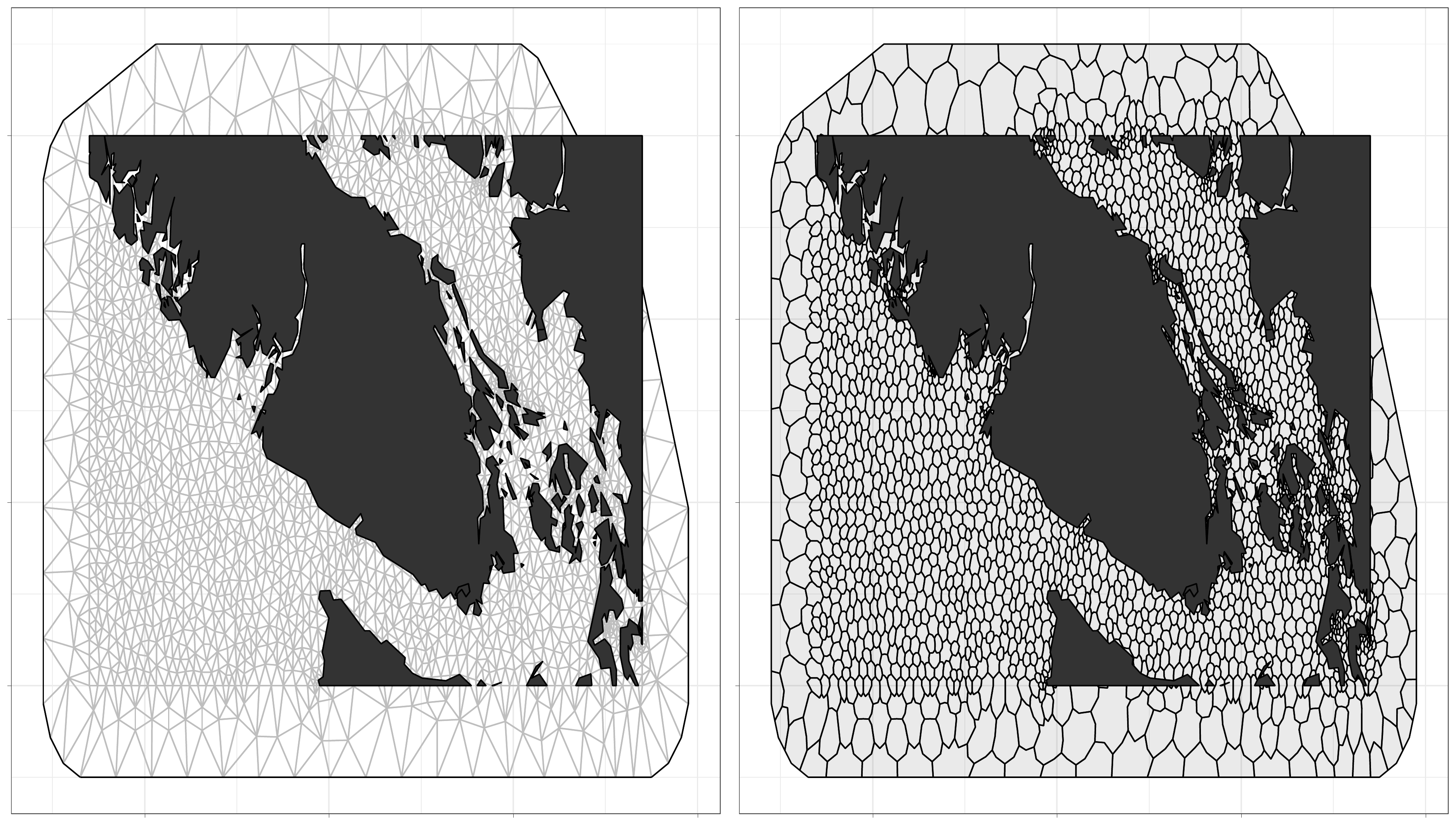}
        \caption{The computational mesh on the left and the corresponding dual mesh on the right, formed by constructing Voronoi polygons around the mesh vertices. The Voronoi polygons form our integration points $A_i$.}
        \label{fig:mesh}
    \end{figure}

Including known or estimated effort from the $O$ observers in the model simply requires evaluating the areal-averaged effort that occurred at each encounter location and around each of the $p$ chosen integration points $\textbf{s}_i : i \in \{1,...,p\}$. We denote the regions around the integration points, $A_j \subset \Omega$. These may correspond to regular lattice cells or as in our example, irregular Voronoi polygons (Fig \ref{fig:mesh}). For the $p$ regions $A_j$, we compute the covariates: 
\begin{align}
    w_{2,o,j} &= |A_j|^{-1} \int_T |\xi_o(t) \bigcap A_j| \textrm{d}t \\
    &= \Tilde{\alpha}_j^{-1} \int_T |\xi_o(t) \bigcap A_j| \textrm{d}t. \nonumber
\end{align}

For the $n$ encounter locations $\textbf{s}_j$, we compute:

\begin{align}
    w_{2,o,j} &= \int_T \mathbb{I} \{ \textbf{s}_j \in \xi_o(t) \} \textrm{d}t.
\end{align}




Often there will be uncertainty surrounding the effort. We present a simple method for accounting for this uncertainty in our later application.


\subsection{Additional details on the results and additional tables}

Environmental covariates were mapped to the integration points $A_i$ and to the sighting locations $\textbf{y}$ for modeling. In cases where we had noisy covariates with missing values, we chose the median covariate value (out of those that spatially-intersect the Voronoi polygon) as the polygon's `representative'. For missing covariates at observation locations, we mapped the non-missing value which was closest in distance to the observation location. Sea-surface temperature (SST) and (log) chlorophyll-A (chl-A) levels were obtained. Monthly chl-A and SST were obtained for each year and averaged over the years. Log transformed covariates were centered to have mean 0 and scaled to have unit variance. Sea surface temperature was not scaled for interpretation reasons.

Next we performed hierarchical centering of our SST and chl-A covariates. This is following the advice of \citet{yuan2017point}, where it was shown that three unique biological insights can be obtained per covariate. In particular, we performed two types of centering: spatial and space-time centering. Centering covariates like this can also improve the predictive performance of models. The 2 hierachical centering schemes applied to both SST and chl-A were compared. We refer to these as covariate sets 1 and 2. 

Models that included a wide range of different latent effects were compared. A unique (sum-to-zero constrained) random walk of second order for each pod was tested, alongside a shared spatial and/or spatio-temporal Gaussian (markov) field across the pods and a unique spatial field for pod L. For the random walk term, we shared the precision parameter across the pods. We put INLA's default logGamma(1, 5e-05) prior distribution on this shared (log) precision. Finally a unique intercept was allowed for each pod. The unique intercepts per pod allow for a different global intensity for each pod to exist across the months, whilst the unique random walk terms per pod allow for a changing relative intensity of each pod across the months. This is chosen based on previous work that found pod J to be the most likely to be present in the Salish Sea year-round \citep{ford2017habitats}. 

We also fitted the models without covariates included in the linear predictor and hence only with spatial and spatio-temporal terms included in the model. We also fitted the models with the covariates kept in, but with the spatial random fields removed. These are inhomogeneous Poisson processes. We did this to attempt to show how the variability seen in the data is captured by covariates and random effects. We also did this to investigate whether or not the spatial distribution of the SRKW intensity (conditioned on the observer effort), changes with month, or whether or not it is spatially static across the months. 

For all spatial fields, we placed PC priors \citep{fuglstad2018constructing} on the GMRF, with a prior probability of 0.01 that the ranges of the fields are less than 15km. We also placed a prior probability of 0.1 that the standard deviations of the fields exceed 3. Thus, our prior beliefs were that the fields are smooth (i.e. the ranges are not too small) and are not too large in amplitude (i.e. the standard deviations are not too large). We did this to reduce the risk of over-fitting the data. The PC priors penalize departures from our prior beliefs under the Occam's razor principle; penalizing models with greater complexity than that specified in our prior.

Now we display the table of coefficients from the `best' model, and the table of DIC values of all tested candidate models. Finally, we display our model-estimated number of sightings per pod and per month, with 95\% credible intervals. We also display the observed number of sightings to check the model's calibration.

\begin{longtable}[c]{|c| c c c c c c|}
\caption{\label{Table:Model8tableMC} A table of posterior estimates of the fixed effects $\boldsymbol{\beta}$, with their 95\% posterior credble intervals for the final Model 8 \textbf{repeatedly fit with 1000 Monte Carlo estimated observer effort fields}. Note that the symbol * denotes `significance' such that the 95\% credible intervals do not cover 0 (no effect), or for the pods represents no difference was found between the relative pod intensities with respect to their 95\% credible intervals. The `change' column displays the change in `significance' of the effect size compared with the results from model 8 \textbf{without the additional MC error from the observer effort}. The `-' symbol denotes no change in significance. None of the directions and hence qualitative conclusions of the effect estimates change.}\\
  \hline
\em & Mean & SD & 0.025 Q & 0.5 Q & 0.975 Q & $\Delta$ \\ 
  \hline
Pod J & -3.84 & 0.71 & -5.23 & -3.83 & -2.44  &  - \\ 
Pod K & -4.57 & 0.71 & -5.95 & -4.56 & -3.20 &  - \\
Pod L & -3.95 & 0.98 & -5.81 & -3.96 & -1.99 &  - \\
  SST month avg & 0.03 & 0.26 & -0.49 & 0.04 & 0.54 & - \\ 
  SST spatial avg$^*$ & -0.37 & 0.17 & -0.70 & -0.37 & -0.05 & - \\ 
  chl-A month avg & 0.31 & 1.11 & -1.89 & 0.26 & 2.58 & -  \\ 
  chl-A spatial avg$^*$ & -1.03 & 0.32 & -1.67 & -1.02 & -0.38 & - \\ 
  SST ST residual$^*$ & -0.67 & 0.05 & -0.77 & -0.67 & -0.57 & - \\
  chl-A ST residual$^*$ & -0.23 & 0.07 & -0.38 & -0.23 & -0.09 & - \\ 
   \hline
\end{longtable}

\newpage

\begin{longtable}[c]{|c c c c c c|}
\caption{\label{Table:DIC} A table showing the DIC values of all the models tested, with the model formulations summarised in the columns. A value of NA implies that model convergence issues occurred.}\\
  \hline
\em Model &\em DIC &\em $\Delta$DIC &\em Covariate Set &\em Shared Field &\em Field for L \\ 
  \hline
0 & 3614 & 5554 & $\times$ & $\times$ & $\times$ \\
1 & 2843 & 4783 & 1 & $\times$ & $\times$ \\
2 & 2707 & 4642 & 2 & $\times$ & $\times$ \\
3 & -1633 & 307 & $\times$ & Spatial & $\times$ \\
4 & -1730 & 210 & $\times$ & Spatio-temporal & $\times$ \\
5 & -1842 & 98 & 1 & Spatial & $\times$ \\
6 & -1851 & 89 & 2 & Spatial & $\times$ \\
7 & -1931 & 9 & 1 & Spatial & Spatial \\
8 & \textbf{-1940} & 0 & 2 & Spatial & Spatial \\
9 & NA & NA & 1 & Spatio-temporal & $\times$ \\
10 & NA & NA & 2 & Spatio-temporal & $\times$ \\
11 & NA & NA & 1 & Spatio-temporal & Spatial \\
12 & NA & NA & 2 & Spatio-temporal & Spatial \\
  \hline
\end{longtable}

\begin{center}
    \begin{figure}
        \centering
        \includegraphics[scale=0.5]{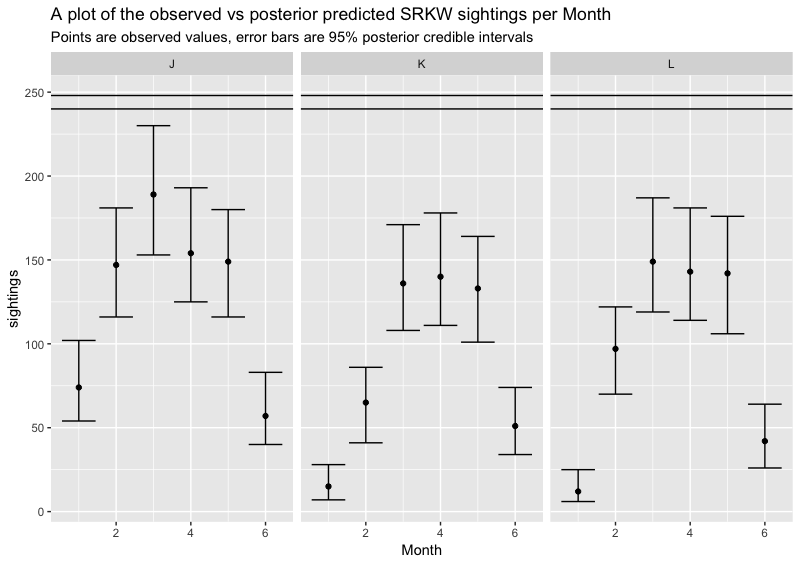}
        \caption{A plot showing the total observed number of sightings made per month with the posterior 95\% credible intervals shown. Results shown are for Model 8 with MC observer effort error. The posterior predictions are made, given the identical observer effort to that estimated for the observed data. Also shown are the horizontal lines showing the maximum possible number of sightings that could be made in months with 30 and 31 days respectively. Posterior credible intervals extending above this upper bound imply the Poisson model is severely misspecified. }
        \label{fig:posteriorpred}
    \end{figure}
\end{center}

\subsection{Pseudo-code for computing the modeling framework in inlabru}

Fitting log-Gaussian Cox process within a Bayesian framework models is greatly simplified with the use of the R package inlabru \citep{inlabru}. Furthermore, inlabru can fit joint models containing many (possibly different) likelihoods, and is able to share parameters and latent effects between them with ease. Numerous other features exist and helper functions are provided to help produce publication-quality plots. For full information and for free tutorials, visit the inlabru website at inlabru.org. The following pseudo-code is largely based on the available tutorials.  

In this section we will demonstrate the simplicity of fitting a joint model to a dataset comprised of a distance sampling survey, and a presence-only dataset with a corresponding observer effort field using inlabru's syntax. For simplicity, suppose we have 1 continuous environmental covariate, called covar1 (e.g. SST), and that it is in the `SpatialGridDataFrame' or `SpatialPixelsDataFrame' class. Next, suppose we have an estimate of the natural logarithm of the observer effort for the presence-only data called logeffort\_po, also of class `SpatialGridDataFrame' or `SpatialPixelsDataFrame'. We assume that the effort took values strictly greater than 0 everywhere before taking the logarithm (or that we have added a small constant to enforce this). 

Suppose we have the observed sighting locations of the individual of interest as two separate objects of class `SpatialPointsDataFrame', one for the survey sightings and one for the presence-only sightings. Call these surv\_points and po\_points and suppose we have thinned the data to ensure any autocorrelation has been removed. Suppose also that we have our transect lines from the survey as an object called surveylines in the class `SpatialLinesDataFrame', and that we know the transect strip half-width (denoted $W$). Finally, suppose that our spatial domain of interest is described by an object called boundary of `SpatialPolygonsDataFrame' class. All `Spatial' objects in the sp package \citep{sp1, sp2} must be in the same coordinate reference system. 

Suppose we wished to estimate a half-normal detection probability function, as a function of distance. To program this in inlabru, we must first define the half-norm detection probability function in R \citep{R}. Let `logsigma' denote the natural logarithm of the standard deviation and `distance' denote the perpendicular distance from the transect to the observed point. Then our function is:

\begin{lstlisting}
halfnorm = function(distance, logsigma){ 
  exp(-0.5*(distance/exp(logsigma))^2)
}.
\end{lstlisting}

Next, given a well constructed Delauney triangulation mesh called `mesh', we construct the spatial random field for the LGCP. Helper functions exist for creating appropriate meshes in inlabru. The code for creating the spatial field, with Matern covariance structure is:

\begin{lstlisting}
matern <- inla.spde2.pcmatern(mesh,
 prior.sigma = c(upper_sigma, prior_probs),
 prior.range = c(lower_range, prior_probr)).
\end{lstlisting}

Here, upper\_sigma, prior\_probs, lower\_range and prior\_probr all define the parameters of the pc prior on the random field \citep{fuglstad2018constructing}. Once again, the tutorials help assist with the choice of prior. Now, we define all the parameters and terms in the model that must be estimated:

\begin{lstlisting}
mod_components <- ~ mySpatialField(map = coordinates, model = matern) +
  beta.covar1(map = covar1, model = `linear') +
  po_search_effort(map = logeffort_po, model = `linear', 
                   mean.linear=1, prec.linear=1e20) +
  logsigma + Intercept_Survey + Intercept_PO.
\end{lstlisting}

Note here that we choose the prior mean and precision of the `po\_search\_effort' field to enforce it to enter the model as an offset. Now we can create the likelihood objects for both data types, each with their own formulae, but sharing components.

\begin{lstlisting}
lik_surv <- like(`cp',
             formula = coordinates ~ Intercept_Survey + mySpatialField +
                       beta.covar1 + log(halfnorm(distance, logsigma)) + 
                       log(1/W),
             data = surv_points,
             components = mod_components,
             samplers = surveylines,
             domain = list(coordinates = mesh))
lik_po <- like(`cp',
             formula = coordinates ~ Intercept_PO + mySpatialField +
                       beta.covar1 + po_search_effort,
             data = po_points,
             components = mod_components,
             samplers = boundary,
             domain = list(coordinates = mesh)
\end{lstlisting}

And then we can fit the joint model and simulate $M$ samples of all of the parameters and latent effects from the posterior distribution.

\begin{lstlisting}
fit_joint <- bru(mod_components, lik_surv, lik_po)
posterior_samples <- generate(fit_joint, n.samples = M).
\end{lstlisting}

Note that once the model object (fit\_joint) is created, the estimated field can be easily plotted, and predictions can easily be made on new datasets and at new locations. Stochastic integration of the field to estimate abundance (for suitable datasets) is also possible using inlabru helper functions. Details can be found in the inlabru tutorials. The above code can scale up to include multiple environmental covariates (including categorical predictors), spatio-temporal fields, and/or temporal effects. Likelihoods of different type (e.g. Bernoulli, Poisson, Gaussian etc.,) can all be included, with this feature becoming especially useful for when the joint estimation of data of differing type is desired.

\subsection{Additional figures}

\subsubsection{Covariate plots}

\FloatBarrier

\subsubsection{Plot of the pod-specific random walk effects}

\FloatBarrier

\subsubsection{Plot of model standard deviation}

\end{document}